\documentclass[aps,prd,twocolumn,floatfix,nofootinbib,showpacs,superscriptaddress]{revtex4-1}

\usepackage{amsmath,amsfonts,amssymb,bm}

\usepackage{graphicx}
\usepackage{color}
\usepackage{subfigure}
\usepackage{appendix}
\usepackage{verbatim}

\definecolor{purple}{rgb}{0.1,0.5,0.4}

\newcommand{\X}{B}
\newcommand{\barrho}{{\bar{\rho}}}
\newcommand{\be}{\begin{eqnarray}}
\newcommand{\ee}{\end{eqnarray}}
\usepackage[colorlinks=true, pdfstartview=FitV, linkcolor=purple, citecolor= purple, urlcolor=blue]{hyperref}

\def \t {\tau}

\def \D {\Delta}

\begin{document}

\title{Stochastic baryon charge transport in relativistic hydrodynamics}

\author{Nicolas Borghini}
\affiliation{Fakult\"at f\"ur Physik, Universit\"at Bielefeld, D-33615 Bielefeld, Germany}

\author{Baochi Fu }
\email{bfu@physik.uni-bielefeld.de}
\affiliation{Fakult\"at f\"ur Physik, Universit\"at Bielefeld, D-33615 Bielefeld, Germany}

\author{Sören Schlichting}
\affiliation{Fakult\"at f\"ur Physik, Universit\"at Bielefeld, D-33615 Bielefeld, Germany}

\date{\today}

\begin{abstract}
We utilize 3+1D stochastic hydrodynamics to study correlations and fluctuations of baryon charge in high-energy heavy-ion collisions. The baryon charge fluctuations are important observables to probe the QCD phase diagram, yet a dynamical description with stochastic hydrodynamics remains challenging due to numerical instabilities and high computational demands. In this work, we employ a linearized approach, allowing us to separately simulate the background energy-momentum evolution of a charge-neutral fluid and the stochastic baryon transport processes, thereby largely reducing computational cost while maintaining sufficient accuracy.
We implement this linearized stochastic charge evolution in the viscous hydrodynamic code MUSIC, and find that it nicely describes the two-point correlation of 1+1D analytical solutions for various equations of state and transport coefficients. 
In particular, the hydrodynamic calculations demonstrate how different rapidity separations probe charge fluctuations originating at different times of the evolution. We also investigate the net baryon correlations after the Cooper--Frye freeze out, which show good consistency with the analytical calculations and indicate that these fluctuation-induced correlations are sensitive to the baryon diffusion coefficient.

\end{abstract}

\maketitle

\section{Introduction}\label{sec:intro}
Relativistic heavy-ion collisions at the Relativistic Heavy Ion Collider (RHIC) and the Large Hadron Collider (LHC) aim at creating deconfined hot and dense QCD matter and to probe QCD phase diagram. 
In high-energy collisions with vanishing baryon chemical potential, lattice QCD calculations show that the phase transition between the hadron gas and the Quark-Gluon Plasma (QGP) is a smooth crossover~\cite{Aoki:2006we,Ding:2015ona}. 
For high baryon number density or equivalently large baryon chemical potential region, various effective models predict a first-order phase transition and suggest the existence of a critical point \cite{Stephanov:1998dy,Pandav:2022xxx,Fukushima:2010bq}. Searching for this critical point and corresponding first-order phase transition line is one of the most prominent goals of the Beam Energy Scan (BES) program at RHIC \cite{Luo:2017faz} and the future experiments at GSI-FAIR, HIAF, J-PARC-HI, and NICA \cite{Bzdak:2019pkr,Almaalol:2022xwv,Ruan:2018fpo}.  

Experimentally, a powerful set of observables are the cumulants of the conserved charges, which in equilibrium can be directly related to thermodynamic susceptibilities~\cite{Asakawa:2015ybt}. 
While experimental measurements at LHC energies~\cite{Braun-Munzinger:2016yjz, ALICE:2019nbs} show remarkable agreement with lattice QCD calculations~\cite{Borsanyi:2011sw,HotQCD:2012fhj, Bazavov:2014xya}, this observation is in fact highly non-trivial, as one is comparing the final state of a non-equilibrium dynamical evolution with a static equilibrium calculation. 
Close to the QCD critical point, conserved charge fluctuations have also been shown to be sensitive to the critical divergence of the correlation length. 
By varying the collision energies, when the thermodynamic parameters of the system at chemical freeze-out line lie in the vicinity the critical point (or critical region for finite volume), models predict that the divergent correlation length results in a non-monotonic behavior of the high-order cumulants \cite{Stephanov:2011pb}.

So far, most of the theoretical predictions for conserved-charge fluctuations assume thermal equilibrium and focus on a system with constant volume and conserved charges~\cite{Karsch:2010ck,Borsanyi:2014ewa,Albright:2015uua,Fu:2016tey,Lu:2022nkz,Almasi:2017bhq,Vovchenko:2017ayq,Bellwied:2019pxh}. 
In contrast, experiments are limited by the detector acceptance, and thus measure particles within a selected kinematic window, and the charges are not exactly conserved. 
Thus a dynamical description is necessary for bridging the gap between equilibrium theory calculations and experiments~\cite{Sakaida:2017rtj, Nahrgang:2018afz, Oliinychenko:2019zfk,Oliinychenko:2020cmr}, see also \cite{Bluhm:2020mpc} for a recent review.

Among the various dynamical models on the market, hybrid hydrodynamics is by now the ``standard model" that connects the bulk properties of QCD matter to the observables of heavy-ion collisions. 
Including stochastic fluctuations in hydrodynamics was first proposed by Landau in the non-relativistic case \cite{landau2013course} and then generalized to the relativistic framework and applied to heavy-ion physics \cite{Kapusta:2011gt}.  
This so-called stochastic hydrodynamics in principle provides a realistic description of the fluctuating observables in heavy-ion collisions, and can be compared apples-to-apples with experimental data. However, because of the large instabilities and high computational demands, the numerical implementation of stochastic hydrodynamics is still challenging. See \cite{Kapusta:2017hfi, Nahrgang:2018afz,Singh:2018dpk,Tang:2023zvj,Nahrgang:2020yxm,An:2020vri, Chattopadhyay:2024jlh,Kuroki:2023ebq} for recent progress and \cite{Bluhm:2020mpc, Basar:2024srd,Wu:2021xgu} for reviews.  

In this paper, we employ linearized stochastic equations to model the fluctuations and transport of baryon charge. This allows us to separately simulate the background energy-momentum evolution and the fluctuating baryon transport. 
We implement this linearized stochastic approach in the 3+1D viscous hydrodynamic code MUSIC, and compute the two-point correlation of the baryon density and the net-proton correlation after Cooper--Frye freeze-out. Our hydrodynamic simulations demonstrate that the longitudinal separation of final-state correlation is determined by the time at which the fluctuation of the conserved currents occurs.    

This paper is organized as follows. In Sec.~\ref{sec:eqs}, we  introduce the linearized stochastic hydrodynamic equations and compute analytical and numerical solutions in 1+1D and for the 3+1D diffusion of baryon number on top of a 1+1D Bjorken flow. 
In Sec.~\ref{sec:model}, we present our implementation of this linear approach in the 3+1D viscous hydrodynamic program MUSIC and investigate its numerical applicability. 
Section~\ref{sec:2pc} presents the two-point longitudinal correlations in MUSIC and the corresponding two-particle correlation is investigated with different input parameters in Sec.~\ref{sec:frz}. 
Finally, Sec.~\ref{sec:sum} briefly summarizes the main results of this paper.

\section{Linear Stochastic hydrodynamics}\label{sec:eqs}

In this section, we introduce the framework of relativistic stochastic hydrodynamics and derive the analytical (numerical) solutions in a Bjorken expanding system. 

\subsection{Hydrodynamic equations with fluctuations}
Relativistic hydrodynamics solves the conservation equations of energy-momentum tensor $T^{\mu\nu}$ and conserved charge currents $N^\mu_i$:  
\begin{equation}
\begin{aligned}\label{eq:consv}
     \partial_{;\mu} T^{\mu\nu} = 0,  \\
     \partial_{;\mu} N^\mu_i = 0,
\end{aligned}    
\end{equation}
where $\partial_;$ defines the covariant derivative and the index $i$ refers to the conserved charges, e.g. baryon number, electric charge, and strangeness. In Landau's approach, the velocity is defined by the energy flow, and thus $T^{\mu\nu}$ and $N_i^\mu$ can be decomposed as
\begin{equation}
\begin{aligned}\label{eq:decompose}
T^{\mu\nu} &= e u^\mu u^\nu - (p + \Pi) \D^{\mu\nu} + \pi^{\mu\nu}, \\
N^\mu_i &=  n_i u^\mu + q_i^\mu,
\end{aligned}
\end{equation}
where $\Delta^{\mu\nu} = g^{\mu\nu} - u^\mu u^\nu$ defines the transverse projector, with the mostly-minus convention for the metric tensor. 
To close the system of equations, we need the equation of state $p(e, n_i)$ and the evolution equations for the dissipative terms $\pi^{\mu\nu}$, $\Pi$ and $q_i^\mu$. In this paper, we neglect the shear and bulk viscous terms and focus on the physics of the baryon charge transport. In the Israel--Stewart formalism, the dissipative current for baryon charge is determined dynamically from the evolution equation \cite{Denicol:2018wdp, Murase:2013tma}:
\begin{equation}\label{eq:diffusion}
\D^{\mu\nu} D_{;} q_\nu = - \frac{1}{\t_q} \left[ q^\mu - \kappa_B \nabla^\mu_{;}\bigg(\frac{\mu_B}{T}\bigg) - \xi^\mu \right],
\end{equation}
where $D_{;} = u^\mu \partial_{;\mu}$ is the covariant time derivative, $\nabla^\mu_{;} = \Delta^{\mu\nu}\partial_{;\nu}$ the transverse derivative, while $\kappa_B$ and $\tau_q$ denote the baryon diffusion constant and the corresponding relaxation time. 
Compared with ordinary fluid dynamics, the most striking difference is the noise term $\xi^\mu$  in this stochastic fluid equation. 
When considering a single dissipative current, using a white noise type $\xi^\mu$ in the differential equation~\eqref{eq:diffusion} was found to be equivalent to a having colored noise on the dissipative current itself \cite{Murase:2019cwc, Hammelmann:2018ath, Kapusta:2017hfi}. While  causality is maintained by a non-vanishing relaxation time $\tau_q$, the same fluctuation-dissipation relation (FDR) as in the Navier--Stokes case applies, and the magnitude of the stochastic white noise is determined by \cite{Kapusta:2011gt}:
\begin{equation}\label{eq:fdr}
    \begin{aligned}
    \langle \xi^\mu (x) \rangle &= 0,  \\
    \langle \xi^\mu (x) \xi^{\nu} (x{'}) \rangle &= 2 \kappa_B \Delta^{\mu\nu} \delta^{(4)}(x-x'),
    \end{aligned}
\end{equation}  
where the angular brackets $\langle ... \rangle $ denotes statistical average: 
The one-point function vanishes and the two-point correlation is proportional to the local baryon diffusion constant $\kappa_B$. In hydrodynamical simulations, the stochastic noise $\xi^\mu$ is sampled event-by-event according to Eq.~(\ref{eq:fdr}) at each time step and evolves as the source terms of the hydrodynamic equations. This stochastic approach is an effective dynamical model that provides a straightforward method to implement fluctuations in the well-established hydrodynamic framework.  
However, for various reasons event-by-event hydrodynamic simulations with sufficient precision are still challenging, and we refer to Sec.~\ref{sec:model} for a more detailed discussion.   

Since we do not consider shear and bulk viscous effects, we also neglect the fluctuations of $T^{\mu\nu}$ and employ a linearized stochastic approach to study the fluctuation and diffusion of the baryon charge in 3+1D hydrodynamics.\footnote{Note that the approach could straightforwardly be extended, by including viscous effects only for the background evolution without any further modifications.} 
We consider a charge neutral fluid at high temperature, i.e., all baryon charge in the system is generated by thermal fluctuations and evolves according to Eqs.~(\ref{eq:consv}) and (\ref{eq:diffusion}). At the leading order, the pressure and temperature are independent of the baryon number density since 
\begin{equation}
    \frac{\partial p}{\partial n_B} \Big|_{n_B = 0} = \frac{\partial T}{\partial n_B} \Big|_{n_B = 0} = 0. 
\end{equation}
As a consequence, the baryon charge fluctuations decouple from the evolution of energy density and flow velocity. 
It allows the temperature and velocity fields to evolve with a zero-chemical-potential equation of state (EoS) as ``background'' on top of which the baryon charge fluctuates and diffuses independently. 
We rewrite this linearized baryon charge conservation and diffusion equations: 
\begin{equation}
\begin{aligned}\label{eq:linear}
     &\partial_{;\mu} (n_B u^\mu + q^\mu) = 0,  \\
     &\D^{\mu\nu} D_{;} q_\nu = - \frac{1}{\t_q} \left[ q^\mu - \kappa_B \nabla^\mu_{;} \bigg(\frac{n_B}{T \chi_B}\bigg) - \xi^\mu \right],
\end{aligned}    
\end{equation}
where we introduced the baryon charge susceptibility $\chi_B = \partial n_B / \partial \mu_B$. In short, the net baryon charge in the system is generated by the stochastic noise term $\xi^\mu$ and diffuses in the deterministic background, with feedback neglected. In the next subsections, we will consider a simple homogeneous condition with Bjorken expansion and find the analytical and numerical solutions in 1+1D and 3+1D respectively. 

\subsection{1+1D analytical solution}

In the Bjorken expansion, we consider a homogeneous temperature profile with flow velocity $u^\mu = \{1, \bm{0}\}$ in Milne coordinates. The linear stochastic equations~(\ref{eq:linear}) can be simplified and rewritten by separating the evolution along the transverse and longitudinal directions as:
\begin{equation}\label{eq:bjorken}
\begin{aligned}
    &\frac{\partial}{\partial \tau} (\tau n_B) = - \tau \bm{\nabla} \cdot \bm{q}, \\
    &\frac{\partial}{\partial \tau} q^\perp = - \frac{1}{\tau_q} \bigg[q^\perp - \kappa_B \partial^\perp \bigg(\frac{n_B}{T \chi_B}\bigg) - \xi^\perp \bigg], \\
    &\frac{\partial}{\partial \tau} q^\eta = - \frac{1}{\tau_q} \bigg[q^\eta - \kappa_B \partial^\eta \bigg( \frac{n_B}{T \chi_B} \bigg) - \xi^\eta \bigg] - \frac{1}{\tau} q^\eta.
\end{aligned}
\end{equation}
Focusing on the 1+1D evolution along the longitudinal direction, Eqs.~(\ref{eq:bjorken}) can be further simplified as:
\begin{align}\label{eq:1d-space}
    \tau_q \frac{\partial^2}{\partial \tau^2} \X &+ \frac{\partial}{\partial\tau} \X - \frac{\kappa_B}{ \tau^2 T \chi_B} \frac{\partial^2}{\partial \eta^2} \X = - \tau \frac{\partial}{\partial \eta} \xi^\eta, 
\end{align}
where we introduced the scaled baryon density
\begin{equation}
\X(x) = \tau n_B(x),
\end{equation}
and assumed that the susceptibility $\chi_B$ is a function of temperature, so that the denominator $T \chi_B$ is homogeneous along the $\eta$ direction and can be factorized out from the derivatives. 
After the Fourier transformation $\int\! f(x)_{} e^{-{\rm i}k_\eta \eta}\,{\rm d}\eta = \tilde{f}(k_\eta)$, the equation in Fourier space is:
\begin{align}\label{eq:1d-fourier}
    \tau_q \frac{\partial^2}{\partial \tau^2} \tilde{\X} + \frac{\partial}{\partial\tau} \tilde{\X} + \frac{D_s}{ \tau^2} k^2_\eta \tilde{\X} = -i\tau k_\eta \tilde{\xi}^\eta, 
\end{align}
where we define the diffusion constant 
\begin{equation}
D_s = \frac{\kappa_B}{T \chi_B}.
\end{equation}
For the solution $\tilde{B}(\tau)$ we make the ansatz:
\begin{equation}
    \tilde{B}(\tau) = \sqrt{\tau}\, e^{-\frac{\tau}{2}} \tilde{f}(\tau).
\end{equation}
Then the homogeneous part of Eq.~(\ref{eq:1d-fourier}) can be simplified by substituting $s = \tau/(2\tau_q)$ and $\nu^2 = D_s k_\eta^2/\tau_q - \frac{1}{4}$:
\begin{equation} \label{eq:1d-bessel}
    s^2 \frac{\partial^2 \tilde{f}}{\partial s^2} + s \frac{\partial \tilde{f}}{\partial s} + (\nu^2 - s^2) \tilde{f} = 0.
\end{equation}
When $D_s$ is a constant and $\nu \in \mathbb{R}$, Eq.~(\ref{eq:1d-bessel}) is the modified Bessel equation with imaginary order. 
The solution is a linear combination of $\tilde{I}_{\nu}(s)$ and $\tilde{K}_{\nu}(s)$, related to the modified Bessel functions via $\tilde{I}_\nu(y) = \Re(I_{{\rm i}\nu}(y))$ and $\tilde{K}_\nu(y) = \Re(K_{{\rm i}\nu}(y))$. 

We define the Green's function $\tilde{G}$ of the homogeneous part of Eq.~\eqref{eq:1d-fourier} such that 
\begin{align} \label{eq:1evo}
    \tilde{B}(k_\eta, \tau) = -{\rm i} k_\eta \int^{\tau}_{\tau_0} \tau' {\rm d}\tau' \tilde{G}(k_\eta; \tau, \tau') \tilde{\xi^\eta}(k_\eta; \tau'). 
\end{align}
In terms of the variable $s$ and $\nu$, this Green's function is given by 
\begin{equation}\label{eq:1d_gf}
    \tilde{G}(k_\eta; s, s_0) = - 2\sqrt{s s_0}\, e^{s_0 - s} \left[ \tilde{I}_{\nu}(s_0 ) \tilde{K}_{\nu}(s) - \tilde{I}_{\nu}(s) \tilde{K}_{\nu}(s_0)  \right]
\end{equation}
where $s_0 = \tau_0 / (2 \tau_q)$ denotes the scaled initial time of the evolution. 
In the case $\tau_q = 0$, the Green's function reduces to the Navier--Stokes form at first order:
\begin{equation}\label{eq:gf-ns}
    \tilde{G}_{\textrm{NS}}(k_\eta; \tau, \tau_0) = e^{D_s k_\eta^2 ( \frac{1}{\tau} - \frac{1}{\tau_0} )}.
\end{equation}
To assess the large-momentum limit, we assume that the finite evolution time satisfies $s \ll \nu$, and use the asymptotic forms of the modified Bessel functions
\begin{equation}\label{eq:bessel-asymp}
\begin{aligned}
    \tilde{I}_\nu(x) &= \sqrt{\frac{\sinh(\pi \nu)}{\pi \nu}} \cos\bigg(\nu \ln\frac{x}{2} -\gamma_\nu\bigg) + \mathcal{O}(x^2) \\
    \tilde{K}_\nu(x) &= -\sqrt{\frac{\pi}{\nu \sinh(\pi \nu)}} \sin\bigg(\nu \ln\frac{x}{2} -\gamma_\nu\bigg) + \mathcal{O}(x^2)
\end{aligned}
\end{equation}
where the phase factor $\gamma_\nu$ satisfies:
\begin{equation}
    \Gamma(1+ {\rm i} \nu) = \sqrt{\frac{\pi \nu}{\sinh(\pi \nu)}}\, e^{{\rm i}\gamma_\nu}. 
\end{equation}
Substituting Eq.~(\ref{eq:bessel-asymp}) into Eq.~(\ref{eq:1d_gf}), the phase factor $\gamma_\nu$ cancels out and we obtain a simple asymptotic function for the large-$k_\eta$ limit:
\begin{equation}\label{eq:green-asymp}
\tilde{G}_{\rm as}(k_\eta; s, s_0) = - 2\sqrt{ s s_0}\, \frac{e^{ s_0 - s}}{\nu} \sin\bigg(\nu\ln\frac{s}{s_0}\bigg). 
\end{equation}
This shows that at sufficiently large momentum (or equally, with a short relaxation time), the response function tends to decay exponentially with the final time $s$, with an oscillating behavior controlled by the ratio of $s$ and $s_0$. 

\begin{figure}
    \centering
    \includegraphics[width=0.48\textwidth, trim= 50 0 0 -10, clip]{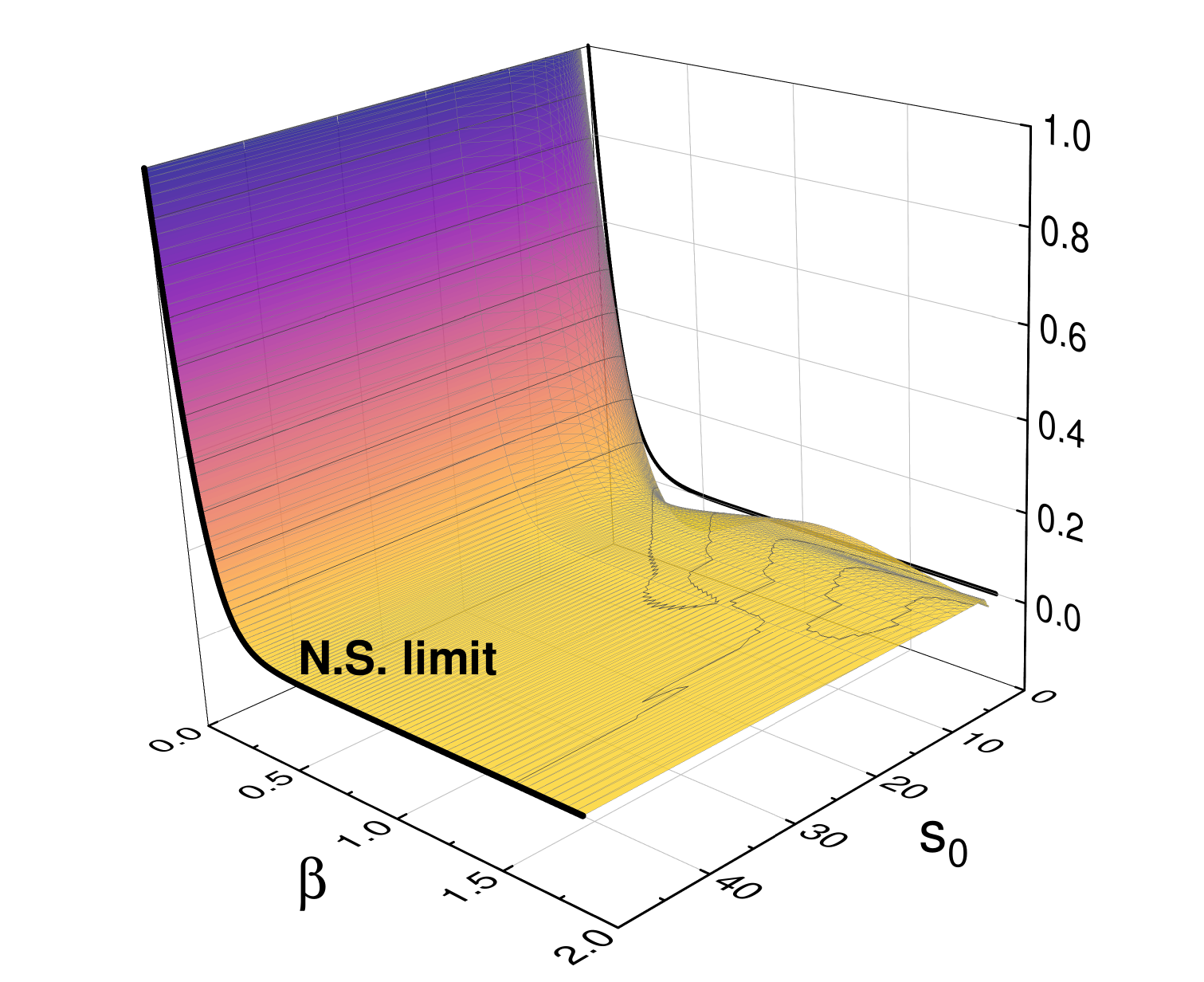}
    \caption{Green's function $\tilde{G}$ for 1+1D baryon charge diffusion in Bjorken expansion, Eq.~\eqref{eq:1d_gf}, as a function of $s_0 = \tau_0 / (2\tau_q)$ and $\beta = D_s k_\eta^2/\tau_F^2$. }
    \label{fig:1+1_GFspec}
\end{figure}

\begin{figure*}
\subfigure{\centering
    \includegraphics[width=0.48\textwidth, trim= 50 0 40 50, clip]{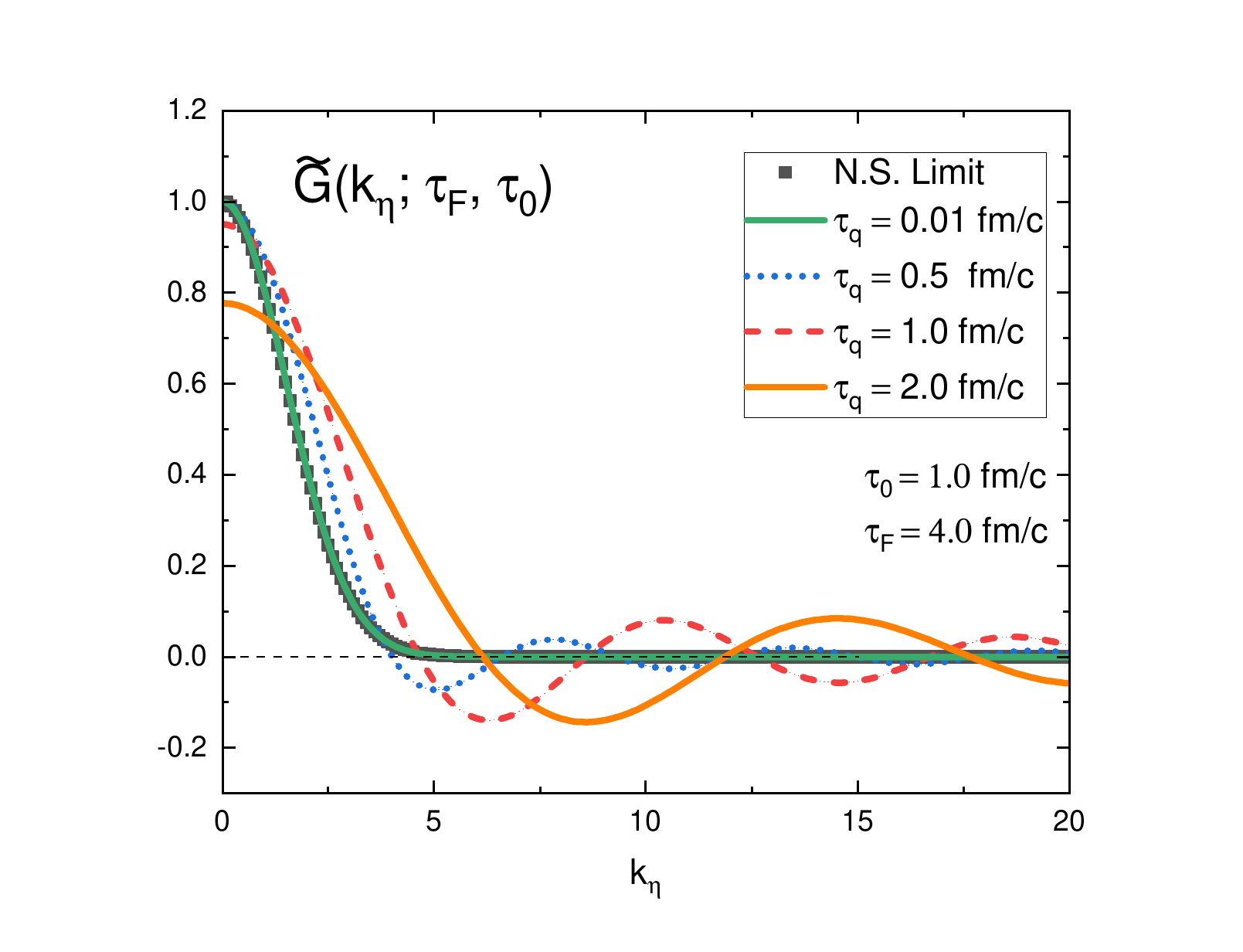}}
\subfigure{\centering
    \includegraphics[width=0.48\textwidth, trim= 70 0 20 50, clip]{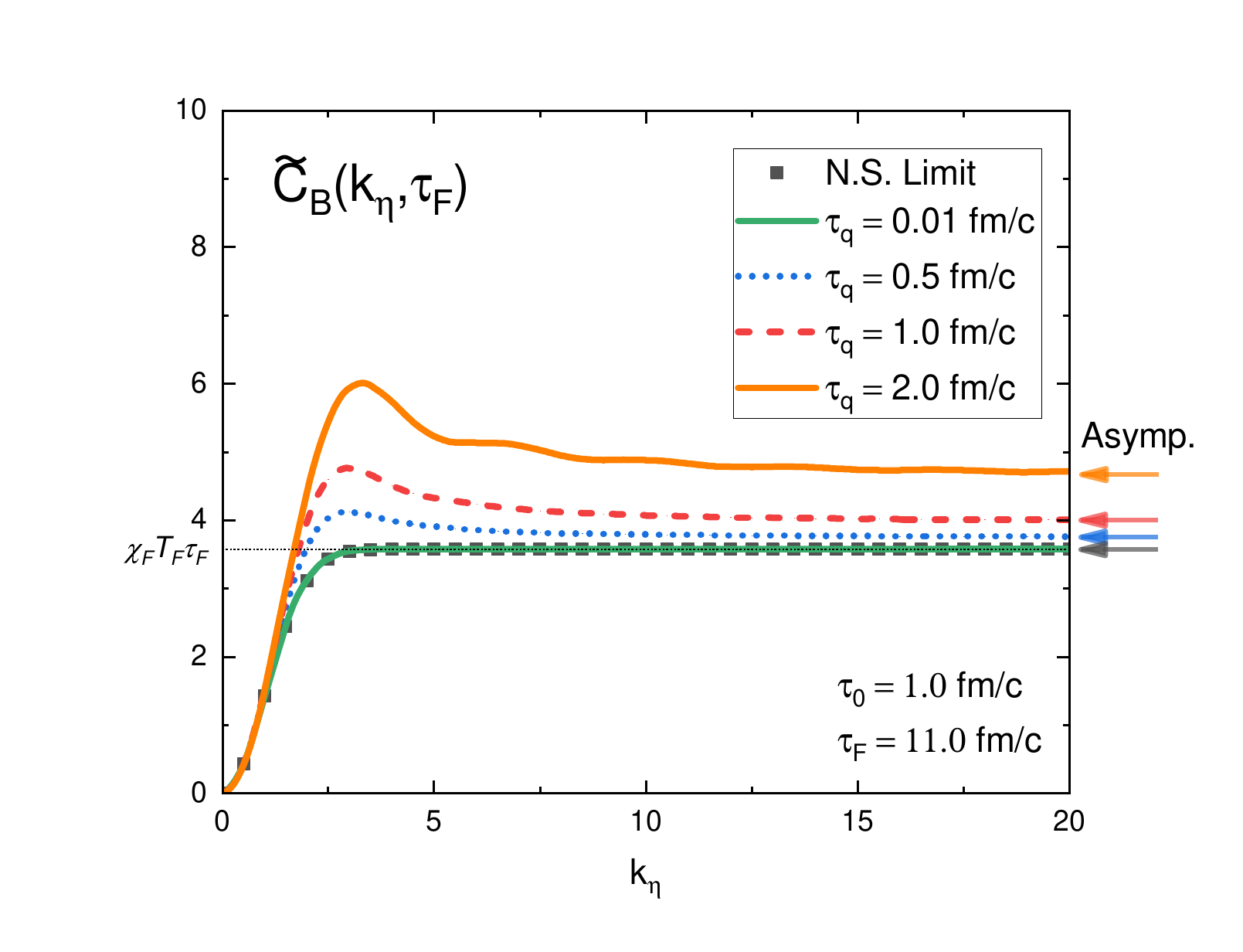}}\vspace{-4mm}
    \caption{The Green's function (left panel) and 2-point correlation (right panel) as functions of longitudinal momentum $k_\eta$ in 1+1D baryon charge diffusion with the Bjorken expansion. The black symbols show the Navier--Stokes limit and the lines show the second-order results with varying relaxation time $\tau_q$. The arrows on the right side of the plot indicate the asymptotic large-$k_\eta$ limit for the corresponding $\tau_q$, as detailed in the text.}
    \label{fig:1+1-2pc}
\end{figure*}

Physically, what can be measured in experiments are the observables on freeze-out surface and we are mostly interested in how they relate to the fluctuations during the hydrodynamical evolution. 
In a 1+1D Bjorken expanding system, freeze out happens at a constant final time $\tau_F$. 
Thus, we fix $\tau = \tau_F$ and investigate how the response function $\tilde{G}(k_\eta; \tau_F, \tau_0)$ changes with the time $\tau_0$ at which the noise is created and with the momentum $k_\eta$. 
Since both the argument $s = \tau/(2\tau_q)$ and the order $\nu$ of the Bessel functions depend on the relaxation time $\tau_q$, we further define \begin{equation}\label{eq:beta_def}
\beta =  \bigg(\nu^2 + \frac{1}{4}\bigg)\frac{\tau_q}{\tau_F^2} = D_s \frac{k_\eta^2}{\tau_F^2},
\end{equation}
which yields the momentum dependence, to disentangle the dependence on $\tau_q$. 

Figure \ref{fig:1+1_GFspec} shows the behavior of the Green's function in the $(\beta, s_0)$ plane. 
By varying $s_0$ at fixed $\beta$, the function probes the final response according to the time at which fluctuations occur (or equally, it controls the relative importance of relaxation). 
In the large-$s_0$ limit, the relaxation time $\tau_q$ approaches zero, and the Green's function is close to the Navier--Stokes baseline $\tilde{G}_{\textrm{NS}}$.
At small $s_0$, the Green's function deviates from the Navier--Stokes result and shows oscillating behavior from non-hydrodynamic modes, which become significant when the evolution time scale is comparable with the relaxation time $\tau_q$. 

Now we derive the two-point correlation in Fourier space. Starting from a general scaled baryon density in 3+1D $B(\tau, x, y, \eta)$, its spatial Fourier transform is:
\begin{equation*}
\tilde{B}(\bm{k},\tau) = \int\! B(\bm{x},\tau) e^{-{\rm i}\bm{k \cdot x}}\,{\rm d}\bm{x}. 
\end{equation*}
We define the associated two-point correlation $\tilde{C}_B(k_\eta, \bf{k_\bot}, \tau)$ in Fourier space as
\begin{equation}\label{eq:1d-cor_gen}
\tilde{C}_B(k_\eta, {\bf k}_\bot, \tau)_{} \equiv \frac{1}{V} \big\langle \tilde{\X}(k_\eta,  {\bf k}_\bot, \tau) \tilde{\X}(-k_\eta,-{\bf k}_{\bot}, \tau) \big\rangle, 
\end{equation}
where the normalization factor $1/V$ is introduced such that the correlation function becomes independent of the spatial volume $V=L_x L_y L_\eta$. 
Indeed evaluating the right-hand side of Eq.~\eqref{eq:1d-cor_gen} yields:
\begin{widetext}
\begin{equation}
\begin{aligned}\label{eq:1d-cor}
       \tilde{C}_B(k_\eta, {\bf k}_\bot, \tau)_{} 
       &= \frac{1}{V}  \int^{\tau}_{\tau_0} \tau_1\, {\rm d}\tau_1  \int^{\tau}_{\tau_0} \tau_2\,{\rm d}\tau_2 (- k_\eta k_\eta' )
       \tilde{G}(k_\eta,{\bf k}_\bot; \tau, \tau_1)
       \tilde{G}(k_\eta',{\bf k}_\bot'; \tau, \tau_2)
       \big\langle \tilde{\xi^\eta}(k_\eta,{\bf k}_\bot; \tau_1) \tilde{\xi^\eta}(k_\eta',{\bf k}_\bot'; \tau_2) \big\rangle\Big|_{\bm{k}'=-\bm{k}} \\
        &= \int^{\tau}_{\tau_0}  {\rm d}\tau_1\,k_\eta^2
       \tilde{G}(k_\eta,{\bf k}_\bot; \tau, \tau_1)
       \tilde{G}(-k_\eta,-{\bf k}_\bot; \tau, \tau_1)
       \frac{2 \kappa_B}{\tau_1}.  \\
\end{aligned}
\end{equation}
\end{widetext} 
%
In the following discussion, we consider a 1+1D correlation:
\begin{equation*}
\tilde{C}_B(k_\eta, \tau) = \tilde{C}_B(k_\eta, {\bf k}_\bot, \tau)_{} |_{\bf{k}_\bot = \bf{0}},
\end{equation*}
and similarly the 1+1D Green's function considered until now is related to the 3+1D Green's function by $\tilde{G}(k_\eta; \tau,\tau_0) = \tilde{G}(k_\eta, \bf{k}_\bot; \tau, \tau_0)|_{\bf{k}_\bot = \bf{0}}$. 
For simplicity, we assume that the baryon diffusion constant scales with the local temperature as $\kappa_B \sim T^3$ \cite{De:2022tkb}, so that its time dependence is $\kappa_B(\tau) = \kappa_B(\tau_0)\, \tau_0/\tau$ for an expansion with an ideal gas EoS. 
When inserting the Navier--Stokes Green's function~(\ref{eq:gf-ns}) in Eq.~(\ref{eq:1d-cor}), the two-point correlation becomes
\begin{align}
    \tilde{C}_{B,\textrm{NS}}(k_\eta, \tau_F)
    &= \!\int^{\tau_F}_{\tau_0} \!\frac{{\rm d}\tau}{\tau^2}  k_\eta^2 
    \tilde{G}_{\textrm{NS}}^2(k_\eta; \tau_F, \tau) 2 \kappa_B(\tau_0)\tau_0 \cr
    &= \chi_F T_F \tau_F \left( 1 - e^{2 D_s k_\eta^2 ( \frac{1}{\tau_F} - \frac{1}{\tau_0} )}  \right).
    \label{eq:1d-corr-ns}
\end{align}
The left panel of Fig.~\ref{fig:1+1-2pc} shows the Green's function $\tilde{G}(k_\eta; \tau_F, \tau_0)$ as a function of the longitudinal momentum $k_\eta$ with $\tau_0 = 1$~fm/$c$ and $\tau_F = 4$~fm/$c$.\footnote{In the following discussions, we mostly focus on a system with freeze-out time $\tau_F = 11~$fm/$c$.  However, here in the Green's function plot we choose $\tau_F = 4~$fm/$c$ to show its structure more clearly, since the details are smeared at late times.} 
The small-$k_\eta$ behavior is close to that found with the first-order Navier--Stokes equations and can be recognized as the hydrodynamic diffusion mode~\cite{Romatschke:2017ejr}. 
At large $k_\eta$, such hydrodynamic diffusion decreases and the non-hydrodynamic mode propagation becomes significant, which leads to the oscillations that become more visible with increasing $\tau_q$. 
Accordingly, changing $\tau_q$ results in a visible ordering of the correlation $\tilde{C}_B(k_\eta, \tau_F)$ plotted in the right panel of Fig.~\ref{fig:1+1-2pc}. 
This ordering can be nicely described by the asymptotic behavior at large $k_\eta$, which yields the values indicated by the arrows on the right side of the plot.
Inserting the asymptotic Green's function $\tilde{G}_{\rm as}(k_\eta; \tau, \tau_0)$ in Eq.~(\ref{eq:1d-cor}), the integration of $\tilde{C}_B$  results in an oscillating function of $s$, which converges to a $k_\eta$-independent value when $k_\eta$ approaches infinity.\footnote{See Appendix \ref{appendix:asymptotic} for more details about the asymptotic solution.}
At small $\tau_q$, the two-point correlation at large $k_\eta$ reproduces the results $\tilde{C}_{B,\textrm{NS}}(k_\eta, \tau_F) = \chi_F T_F \tau_F$ of the Navier--Stokes theory as expected. 

\subsection{3+1D numerical solution}

We next turn to the 3+1D diffusion of baryon number in a transversely homogeneous fluid undergoing Bjorken expansion. Since we are only interested in charge density fluctuations, we can decouple the dynamics of the transverse components of the transverse currents, such that at linear order, the stochastic hydrodynamic equations~(\ref{eq:bjorken}) in Fourier space take the form

\begin{equation}\label{eq:matrixk}
    \frac{\partial}{\partial \tau} \tilde{Q}(k) - \tilde{M}(k) \tilde{Q}(k) = \tilde{\Xi}(k) = 
    \begin{pmatrix}
        0 \\
        \tilde{\xi}^\perp / \tau_q \\
        \tilde{\xi}^\eta / \tau_q \\
    \end{pmatrix}.
\end{equation}
with $ Q = \left( \X, q^\perp, q^\eta \right)^T$, where
$q^{\perp}={\bf k}_{\bot}\cdot{\bf q}_\bot / |k_\bot|$ and $\xi^{\perp}= {\bf k}_{\bot}\cdot\bm{\xi}^{\perp}/|k_\bot|$ are the longitudinal projections of the transverse current and noise, respectively.
Expressing $D_s=\kappa_B / (T\chi_B)$ as before, the matrix $\tilde{M}(k)$ reads 
\begin{equation}\label{eq:M(k)_3+1D}
    \tilde{M}(k) = - 
    \begin{pmatrix}
    0 & i k_\perp \tau & i k_\eta \tau \\[1.5ex]
    i k_\perp \dfrac{D_s}{\tau_q \tau}  & \dfrac{1}{\tau_q} & 0\\[2ex]
    i k_\eta  \dfrac{D_s}{\tau_q \tau^3} & 0 & \dfrac{1}{\tau_q} + \dfrac{1}{\tau}\\[2ex]
    \end{pmatrix}.
\end{equation}
\begin{figure*}
\subfigure{\centering
    \includegraphics[width=0.48\textwidth, trim= 50 0 50 0, clip]{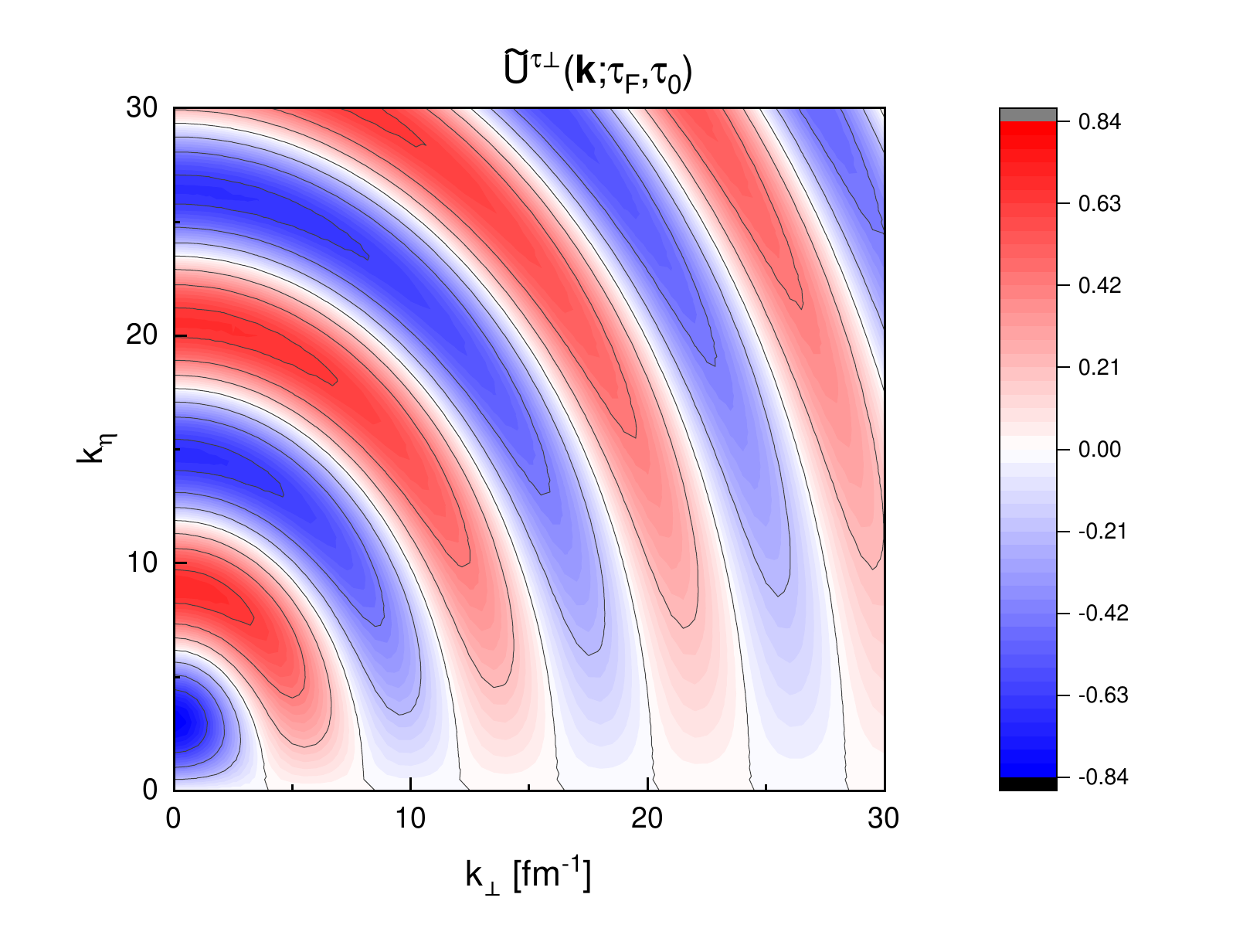}}
\subfigure{\centering
    \includegraphics[width=0.48\textwidth, trim= 50 0 50 0, clip]{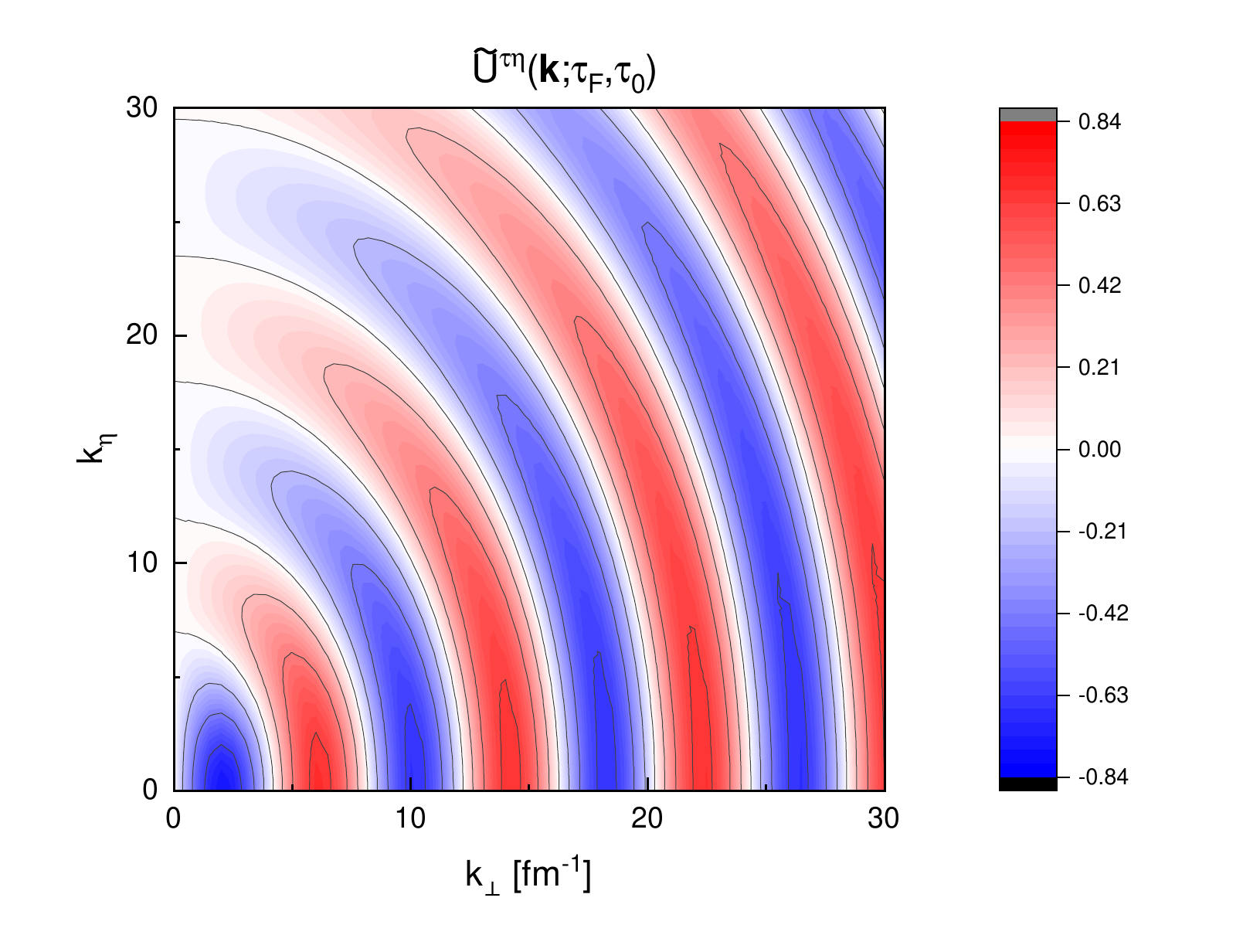}}\vspace{-4mm}
    \caption{The component $\tilde{U}^{\tau \bot}$ (left panel) and $\tilde{U}^{\tau \eta}$ (right panel) on the $(k_\perp, k_\eta)$ plane, respectively. The initial time is $\tau_0 = 1$~fm/c and the final time $\tau_F = 4$~fm/c.  }
    \label{fig:2d_gf}
\end{figure*}

\begin{figure}
    \centering
    \includegraphics[width=0.45\textwidth, trim= 0 0 0 0, clip]{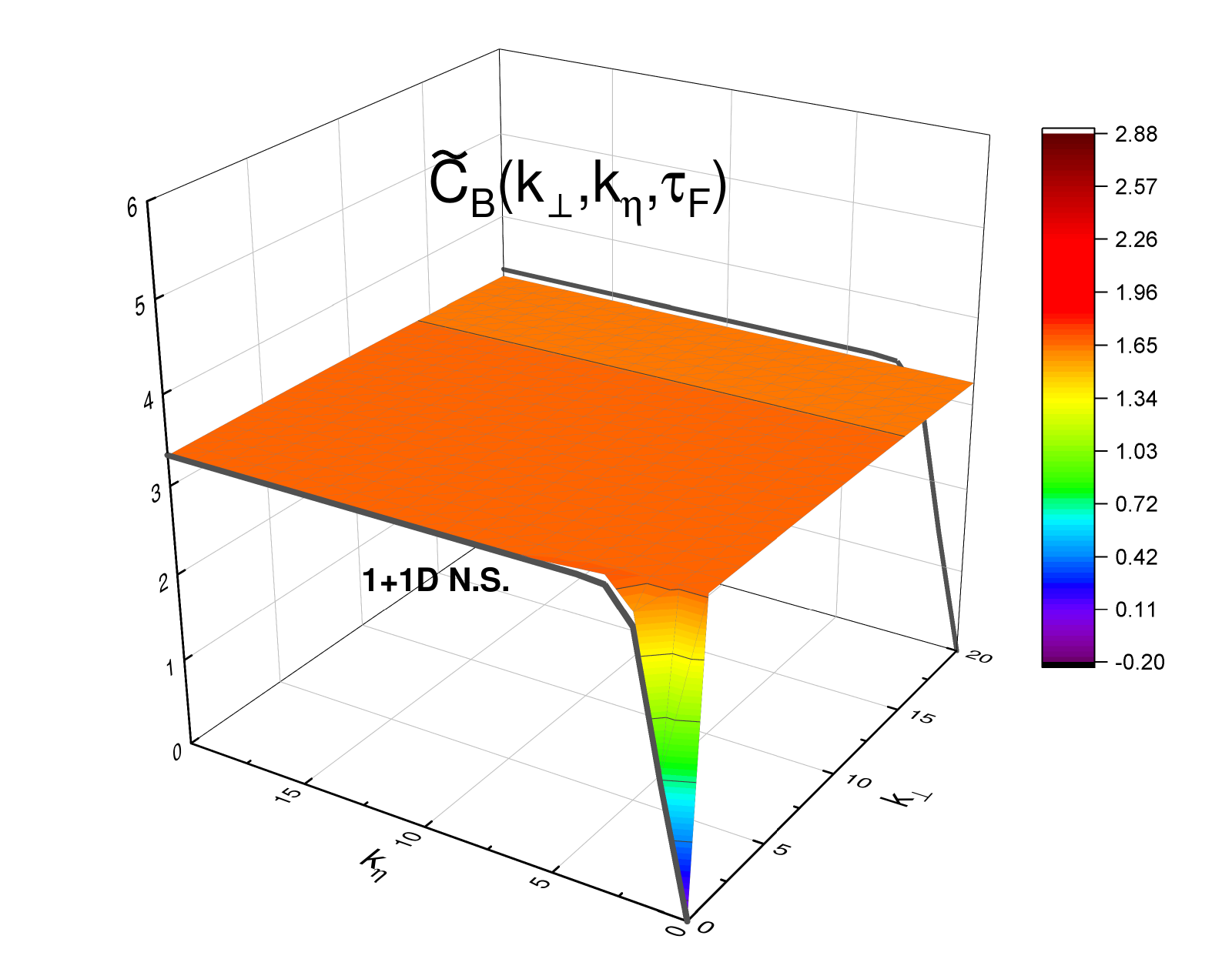}
    \caption{The two-point correlation in 3+1D baryon-number diffusion in the $(k_\perp, k_\eta)$-plane with $\tau_q = 0.001$~fm/$c$ and final time $\tau_F = 4$~fm/$c$. The black line at $k_\perp=0$ is the Navier--Stokes result in 1+1D longitudinal evolution.}
    \label{fig:2d_2pc}
\end{figure}

The linear system of equations~(\ref{eq:matrixk}) can be solved numerically, with the formal solution $\tilde{Q}(\t) = \tilde{U}(\tau, \tau_0) \tilde{Q}(\tau_0)$. At the leading order, the matrix entries $\tilde{U}^{\tau \perp}$ and $\tilde{U}^{\tau \eta}$ represent the 1+1D transverse and longitudinal Green's functions:
\begin{align}
\tilde{\X}(\bm{k}, \t) = &\int_{\tau_0}^{\tau} {\rm d}\tau'\, \tilde{U}^{\tau \perp}(\bm{k}; \tau, \tau') \tilde{\xi}^\perp(\tau') /\tau_q \cr
    &+\int_{\tau_0}^{\tau} {\rm d}\tau'\, \tilde{U}^{\tau \eta}(\bm{k}; \tau, \tau') \tilde{\xi}^\eta(\tau') /\tau_q. 
\end{align}
Figure \ref{fig:2d_gf} displays these two components in the $(k_\perp, k_\eta)$-plane.  
They show the same damping trend along the transverse or longitudinal direction as observed in the 1+1D case, and oscillate along the other direction. When the relaxation time is such that $\tau_q \ll \tau$, the additional term $1/\tau$ in the third line of Eq.~\eqref{eq:M(k)_3+1D} will not change the structure significantly and a rough symmetry between $\tilde{U}^{\tau \perp}$ and $\tilde{U}^{\tau \eta}$ can be found in Fig.~\ref{fig:2d_gf}.

The resulting 3+1D correlation $\tilde{C}_B(\bm{k}, \tau_F)$ is shown in Fig.~\ref{fig:2d_2pc} for a very small $\tau_q = 0.001$~fm/$c$. 
In the zero transverse mode $k_\perp = 0$, the correlation matches the 1+1D Navier--Stokes limit. 
At large $\bm{k}$, the correlation tends to an asymptotic value, which can be taken as a generalization of the 1+1D case in Fig.~\ref{fig:1+1-2pc}.

\section{Implementation in MUSIC}\label{sec:model}

After having derived the analytical and numerical solutions for a system undergoing Bjorken expansion, we now further apply the linear stochastic equations~(\ref{eq:linear}) in more realistic hydrodynamic simulations. 
We extend the 3+1D viscous hydrodynamic code MUSIC with an implementation of stochastic baryon charge transport. 
In this section, we introduce this numerical scheme and perform verifications with analytical solutions to test its applicability. 

\subsection{Numerical Scheme}

MUSIC is a widely used 3+1D program that solves the hydrodynamic conservation equations and the deterministic equations of motions for the dissipative terms \cite{Schenke:2010nt, Paquet:2015lta, Denicol:2018wdp}. In this work, we extend the ordinary MUSIC code to include the stochastic baryon-charge fluctuations $\xi^\mu$ in the diffusion equation Eq.~(\ref{eq:diffusion}). From the FDR, the mean value of the noise term vanishes and the variance is proportional to a Dirac distribution. 
In Milne coordinates, the latter includes an additional Jacobian factor $\sqrt{-g} = \tau$~\cite{Murase:2015oie}:
\begin{equation}
    \delta^{(4)} (x - x_0) \rightarrow \frac{1}{\sqrt{-g}} \prod_{i} \delta(x_i - x_{i,0}) 
\end{equation}
In numerical simulations the one-dimensional Dirac distributions on the right hand side are defined discretely as $\delta(x_i - x_{i,0})|_{x_i = x_{i,0}} = 1/\Delta x_i$, with $\Delta x_i$ the grid spacing in the $x_i$-direction.
Accordingly the covariance of the input white noise~\eqref{eq:fdr} is given by:
\begin{equation}\label{eq:noise-num}
    \langle \xi^\mu_{x_0} \xi^\nu_{x_0} \rangle = \frac{2 \kappa_B \Delta^{\mu\nu}}{ \tau \Delta \tau \Delta x \Delta y \Delta \eta}. 
\end{equation}
The magnitude of the noise is therefore determined by the grid-cell size, so that large gradients challenge the algorithm and may lead to instability, as will be discussed in more detail in next subsection.  
For convenience, we only consider the spatial noise diffusion and deduce the time component $q^\tau$ of the dissipative baryon-number current from the transversality condition $q^\mu u_\mu = 0$. 

\begin{figure*}[hbt]
    \centering
    \includegraphics[width=0.78\textwidth, trim= 50 0 100 0, clip]{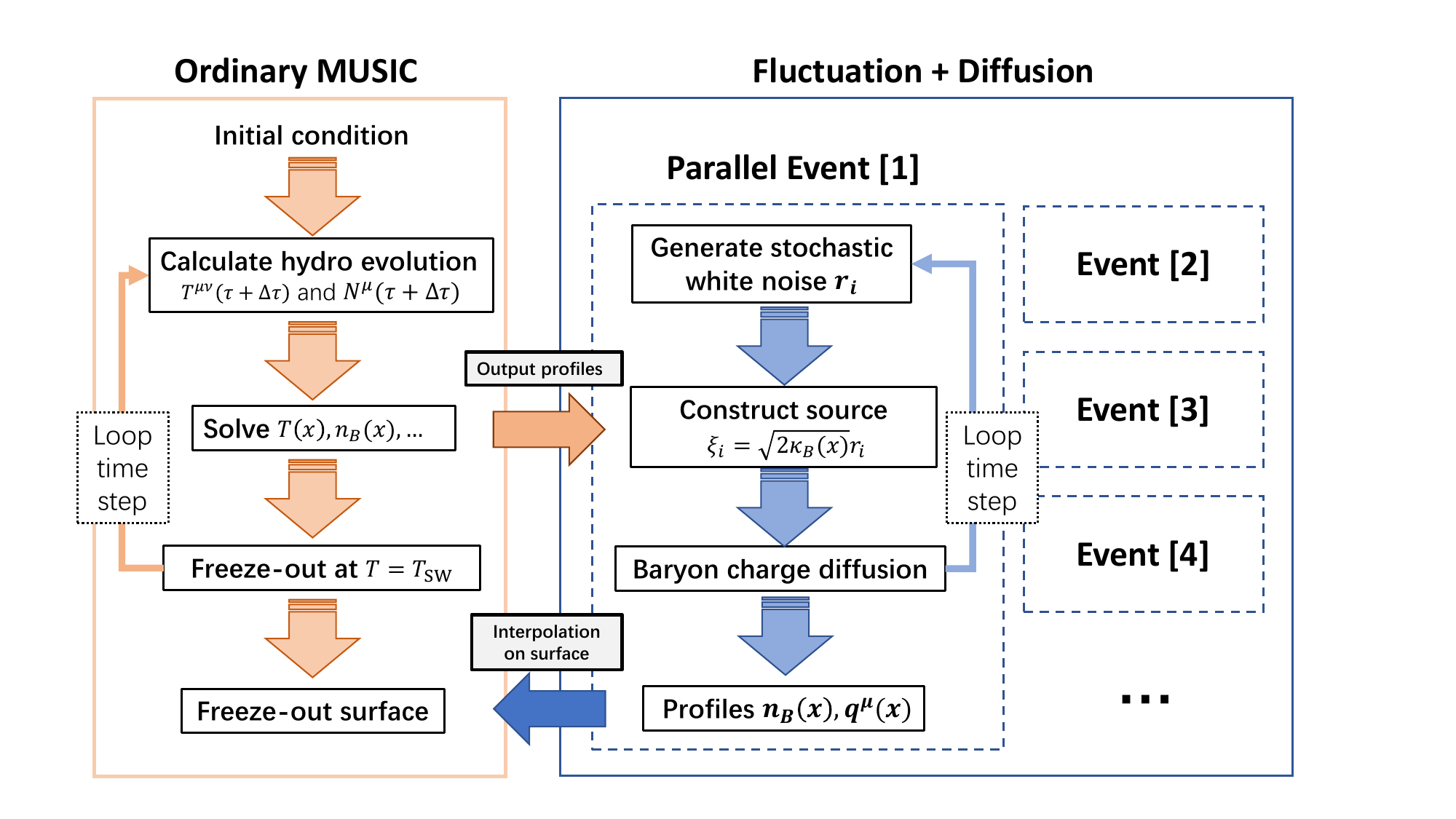}
    \caption{The flow chart of MUSIC with fluctuating baryon-number diffusion events.}
    \label{fig:chart}
\end{figure*}

As the linearized approach allows us to evolve the deterministic $T^{\mu\nu}$ and the fluctuation of $N^\mu$ independently, we can run multiple fluctuating baryon-diffusion events in parallel on top of a single charge-neutral hydrodynamic background. 
Figure \ref{fig:chart} describes this procedure and we briefly summarize the procedures as follows:
\begin{itemize} 
    \item The numerical simulation is divided in two parts: ordinary hydrodynamics and fluctuating events. For the ordinary part, we run MUSIC with smooth initial conditions with zero chemical potential. The temperature and velocity profiles are output at every time step until freeze-out. 
    \item For each fluctuating event, we generate the stochastic noise term $\xi^\mu$ according to Eq.~(\ref{eq:fdr}) and calculate the evolution of the resulting current by Eq.~(\ref{eq:linear}) step by step, with the fluid velocity and temperature evolution provided by the background. 
    \item When the system reaches the freeze-out condition, we calculate the interpolated values of the fluctuating $n_B$ and $\mu_B$ at the center of the freeze-out hypersurface patches and output them for further analysis. 
\end{itemize}

\begin{figure}[ht]
    \centering
    \includegraphics[width=0.48\textwidth, trim= 0 0 0 0, clip]{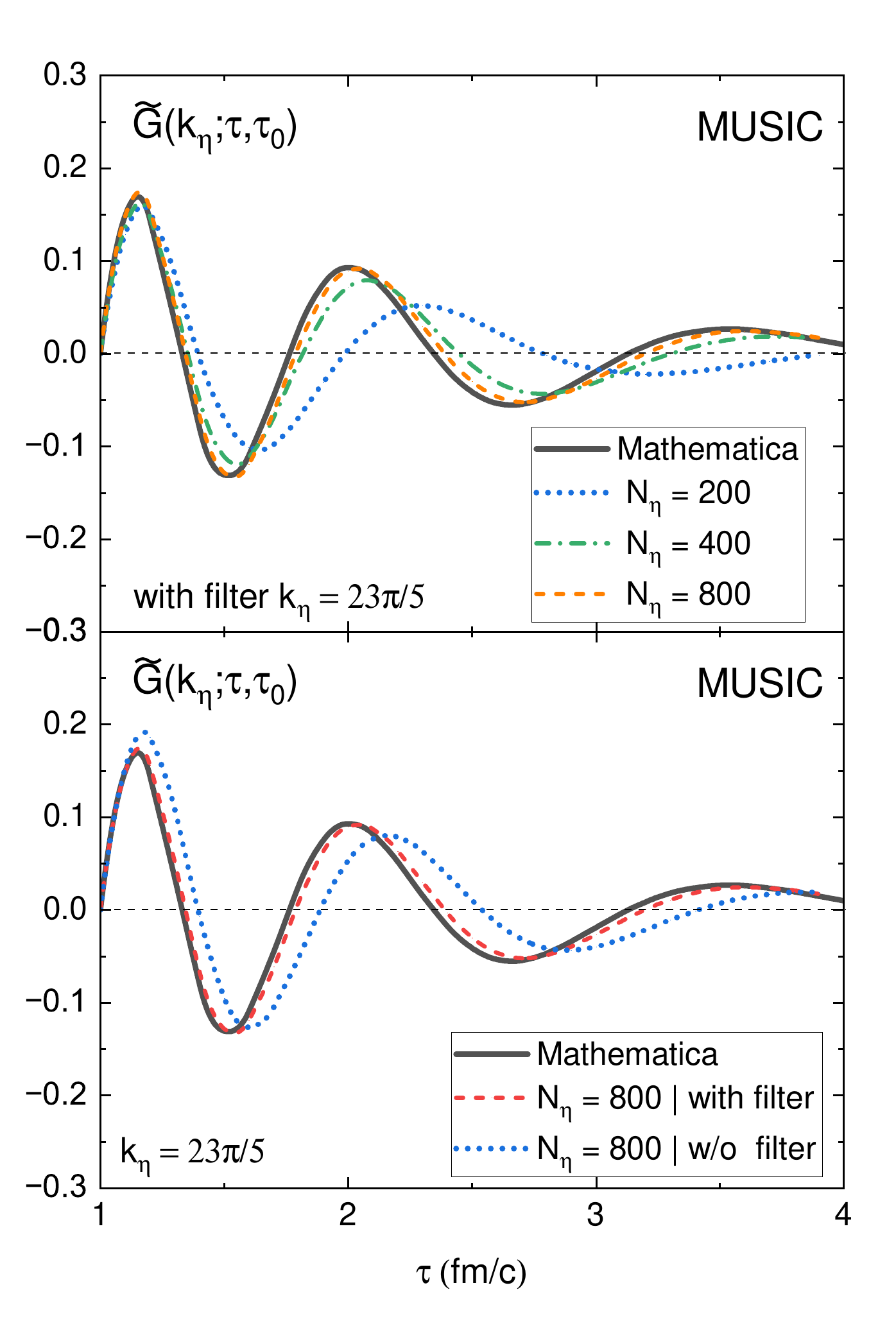}\vspace{-4mm}
    \caption{Dependence of the Green's function extracted from MUSIC on the grid size (upper panel) and influence of nonlinear effects (lower panel). When the frequency filter is applied, all higher modes are set to zero. The spacetime rapidity range is $(-5,5)$, corresponds to a grid spacing $\Delta \eta_s = 0.05$ when $N_\eta = 200$ and $\Delta \eta_s = 0.0125$ when $N_\eta = 800$.}
    \label{fig:gf_size}
\end{figure}
\begin{figure}
    \centering
    \includegraphics[width=0.48\textwidth, trim= 20 0 20 0, clip]{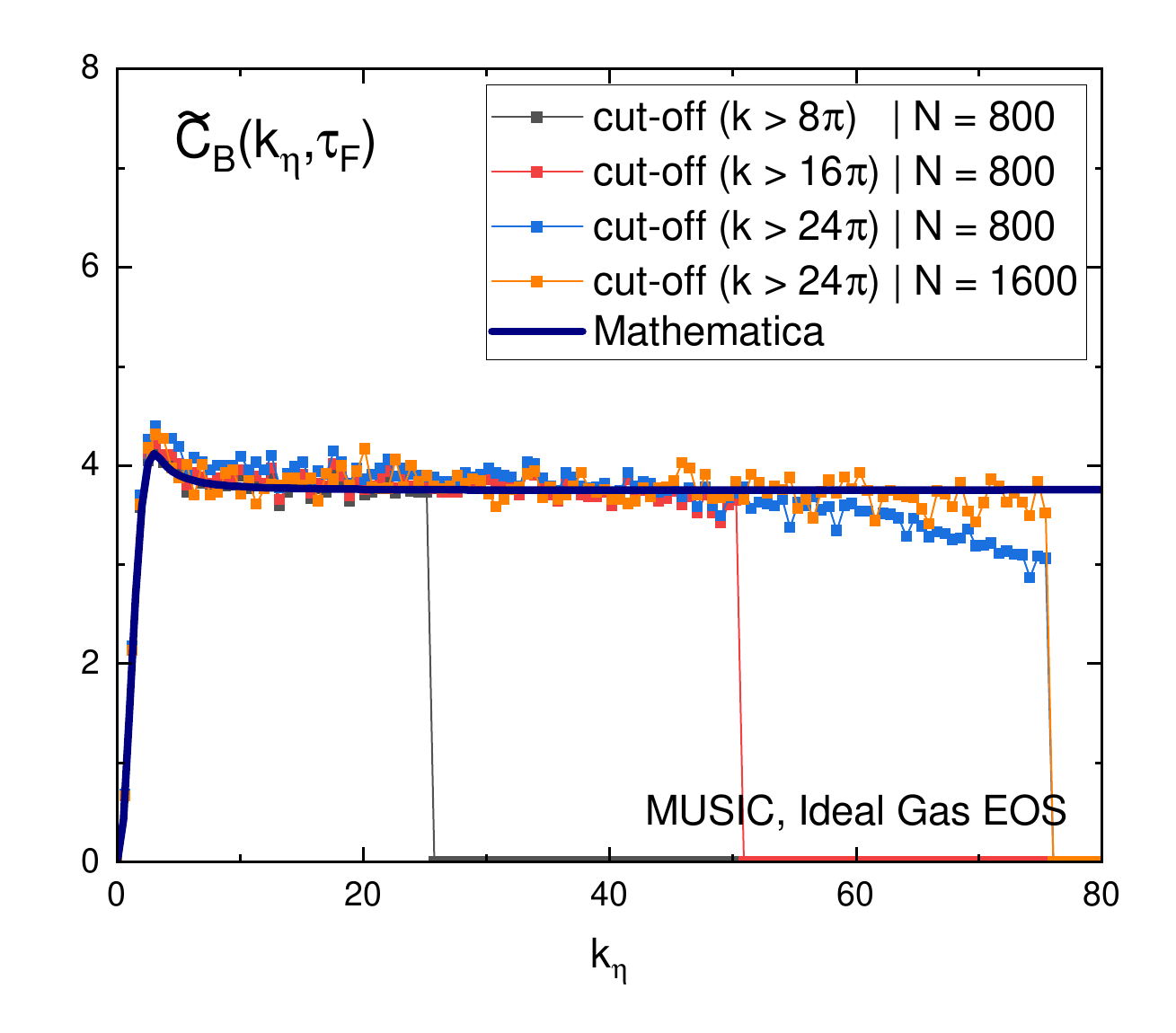}\vspace{-4mm}
    \caption{Two-point correlation as a function of $k_\eta$. The colored lines show different cut-off momentum and grid size combination in MUSIC simulation. }
    \label{fig:2pc_cutoff}
\end{figure}
With this numerical scheme, we can now compute the linearized stochastic baryon-charge fluctuation and transport event-by-event, which largely accelerates the fluctuating hydrodynamic simulation. 
However, due to the Dirac-delta form of the white noise, the stochastic process depends on the grid size, and large gradients are still a non-trivial challenge for the numerical simulation. Before we present the MUSIC results, we will discuss the applicability of this numerical framework and test its accuracy.

\subsection{Numerical applicability}

Stochastic fluid dynamics, also known as stochastic Langevin-hydrodynamics, is developed from the ordinary deterministic relativistic hydrodynamic models that have been widely used in heavy-ion physics. However, in contrast to ordinary hydrodynamics, there are significant challenges in realizing accurate numerical calculations of such a stochastic model~\cite{Bluhm:2020mpc,Basar:2024srd},
which remains an active field of research~\cite{Bhambure:2024gnf,Chattopadhyay:2024jlh,Chattopadhyay:2024bcv}. 

Generally, in ordinary hydrodynamic simulations, the precision of the numerical solution of the partial differential equations can be improved by refining the spatial and temporal resolution of the discretized grid. 
However, in the presence of stochastic noise, the resolution cannot be improved naively by shrinking the spatial grid size (or time step), since the noise term is inversely proportional to the volume size, and thus will eventually diverge as the grid size becomes smaller. 
Stated differently, the noise persists down to the smallest scales irrespective of the discretization and may lead to numerical instabilities or unphysical behavior due to the large gradients. 
Moreover, if fluctuations on all scales are considered, the system can deviate significantly from the local equilibrium condition and make the assumption underlying a hydrodynamic description invalid \cite{Basar:2024srd,Murase:2015oie}.
We can circumvent these issues, by introducing a cut-off scale $l_{\rm filter}$ for the noise, that decouples the scale of thermodynamic fluctuations from the grid discretization $l_{\rm grid}$. We then expect to obtain a faithful numerical solution of the linearized stochastic differential equation, as long as
$$l_{\rm grid} \ll l_{\rm filter}$$

\begin{figure*}
    \centering
    \includegraphics[width=0.65\textwidth, trim= 0 0 0 0, clip]{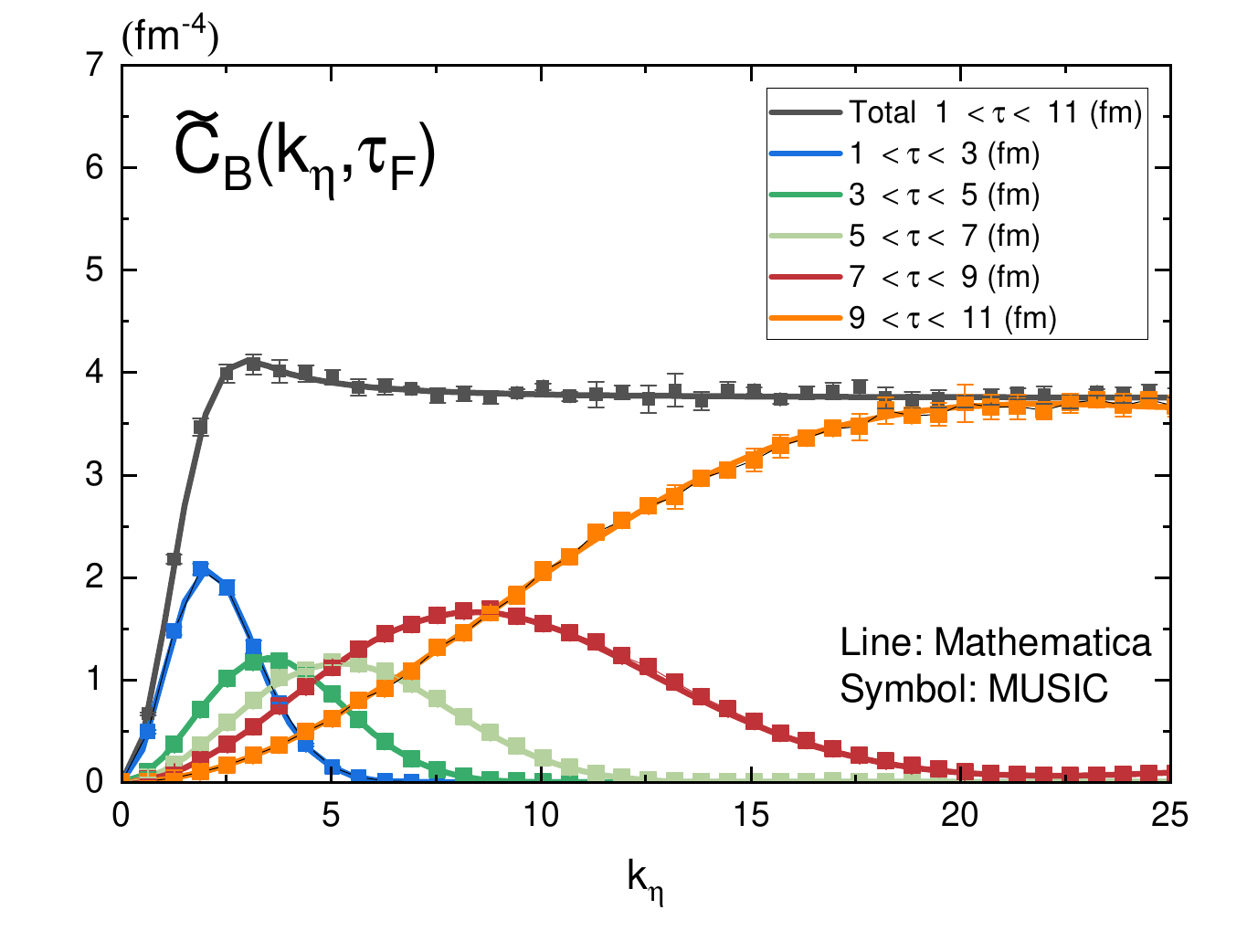}\vspace{-4mm}
    \caption{Two-point correlation as a function of longitudinal momentum $k_\eta$ from Mathematica analytical calculation and MUSIC simulations, respectively. The colored lines stand for the different time intervals at which the noise originates.}
    \label{fig:2pc_music}
\end{figure*}

Besides avoiding large-gradient instabilities, a noise filter is also necessary to eliminate nonlinear effects introduced by the flux limiter. In MUSIC, the Kurganov--Tadmor algorithm \cite{Kurganov:2000ovy} applies a flux limiter that automatically switches the local derivative to avoid spurious oscillations: 
\begin{equation}
(\rho_x)_j =
\hbox{minmod}
\bigg(
\theta\frac{\barrho_{j+1}-\barrho_j}{\Delta x}, 
\frac{\barrho_{j+1}-\barrho_{j-1}}{2\Delta x},
\theta \frac{\barrho_{j}-\barrho_{j-1}}{\Delta x}
\bigg)\nonumber
\end{equation}
with $\theta$ a parameter between 1 and 2, and  
\begin{equation}
\hbox{minmod}(x_1, x_2, \cdots) = \left\{
\begin{array}{ll}
\hbox{min}_j\{x_j\},& \hbox{if $x_j> 0$ $\forall j$}\\
\hbox{max}_j\{x_j\},& \hbox{if $x_j< 0$ $\forall j$}\\
0, & \hbox{otherwise.}\nonumber
\end{array}
\right.
\end{equation}
However, this ``minmod" function will inevitably lead to nonlinear effects such that the evolution of the low-momentum modes is affected by the high-momentum modes even when solving linearized equations. 
To eliminate such nonlinear effects and the resulting instabilities, here we introduce a momentum cut-off $k_{\rm cut}$ and filter out the higher-frequency modes of the noise, setting them to zero by hand. 
This limits the noise resolution to scales larger than the filter scale. 

To evaluate the grid size dependence and the nonlinear effects, we consider the hydrodynamic response to the initial noise $\xi^\mu(\tau_0, x)$ and neglect the noise originating during the rest of the evolution. From Eq.~(\ref{eq:1d_gf}), the final baryon number profile is related to the input initial noise with the help of the Green's function:
\begin{align} \label{eq:gf_extract}
    \tilde{\X}(k_\eta, \tau) = -{\rm i} k_\eta \tau_0 \Delta\tau\, \tilde{G}(k_\eta; \tau_0, \tau) \tilde{\xi^\eta}(k_\eta; \tau_0). 
\end{align}
With this relation, we are able to extract the Green's function from the MUSIC simulation and compare it with the analytical solutions to test the numerical precision. 

The upper panel of Fig.~\ref{fig:gf_size} shows the grid size dependence of the Green's function with a relatively large momentum cut-off $k_\eta = 23 \pi /5$ in MUSIC with frequency filter. 
Here we set the longitudinal size of the system to $L_\eta = 10$ and vary the number $N_\eta$ of grid points in $\eta_s$ to obtain different grid steps $\Delta \eta_s$. 
With a typical hydrodynamical setting $N_\eta = 200$, i.e.\ $\Delta \eta_s = 0.05$, the extracted Green's function deviates from the mathematical benchmark and is smeared out by the numerical viscosity. 
Such deviation is suppressed by reducing the grid size and we find that the exact result is well reproduced with $N_\eta = 800$ ($\Delta\eta_s = 0.0125$) with an acceptable numerical precision. 
The lower panel of Fig.~\ref{fig:gf_size} shows the influence of nonlinear effects on the Green's function extracted in MUSIC. With sufficiently small grid size, the result with a frequency filter above $k_\eta = 23 \pi/5$ describes the Green's function. 
Whereas without filter, the higher-frequency modes lead to a sizable phase shift of the Green's function, which would affect the results of a realistic hydrodynamic simulation if all the modes were combined. 

Figure \ref{fig:2pc_cutoff} further shows the two-point correlation in MUSIC with different cut-off momenta and grid sizes. 
This two-point correlation characterizes the hydrodynamic fluctuation and most importantly, it can be measured with final particles. 
Although nonlinear effects and the finite grid size induce deviations of the Green's function from its exact value, the MUSIC values match the two-point correlation at low momenta, and the results are insensitive to the cut-off momentum or grid size. 
This consistency demonstrates that the implementation of the linearized stochastic hydrodynamics in MUSIC is working and can be used to describe the two-point correlation within the required precision. 
In the following calculations, we set $N_\eta = 800$ and cut-off momentum $k_\eta = 8 \pi$. 

\section{Two-point correlations in MUSIC}\label{sec:2pc}

Now that we have established the consistency of our numerical calculations, we move on to calculate conserved charge correlations on the freeze-out surface for a transversely homogeneous system undergoing Bjorken flow. 
If not stated otherwise, we employ an ideal gas equation of state $p = \frac{1}{3} e$, such that the time-dependence of temperature follows a power law $T \sim \tau^{1/3}$. 
This choice allows us to directly compare our MUSIC results with 1+1D analytical calculations. 
For the chemical potential, we assume $\chi_B \propto T^2$ \cite{De:2022tkb} and $\kappa_B \propto T^3$, which result in a constant diffusion coefficient $D_s = 0.28$~fm.  
We also set a constant relaxation time $\tau_q = 0.5$~fm/$c$ for our comparisons. 

Figure~\ref{fig:2pc_music} is one of the main results in this paper, and compares the two-point correlation in MUSIC with 1+1D Mathematica results in a longitudinally expanding system. 
To evaluate how the final observables are affected by the time at which the fluctuations originate, here we consider different intervals for this origin: for instance, the line ``$1 < \tau < 3$~(fm)" represents the two-point correlation $\tilde{C}_B(k_\eta; \tau_F)$ at $\tau_F = 11~$fm/$c$ with noise created at $1 < \tau < 3 $~fm/$c$, and neglect the fluctuations arising in the rest of the evolution.  
In event-by-event simulations, the error bars denote the statistical error calculated by a Jackknife method in 10,000 fluctuating events.
We find that the hydrodynamic calculation nicely matches the analytical results from Mathematica and also reproduces the dependence on the time interval at which fluctuations appear. 
The low momentum part, corresponding to long range correlation in coordinate space, is dominated by the fluctuations generated at early times. 
This is readily understood, since long range correlations require a longer time to diffuse the fluctuation signal. 
In contrast, the contribution of late-stage fluctuations (as would be the case of the possible critical fluctuations in simulations at finite baryon chemical potential) are relatively more important in the short-range correlations, i.e.\ at large momentum. 

\begin{figure*}
\subfigure{\centering
    \includegraphics[width=0.49\textwidth, trim= 80 10 45 62, clip]{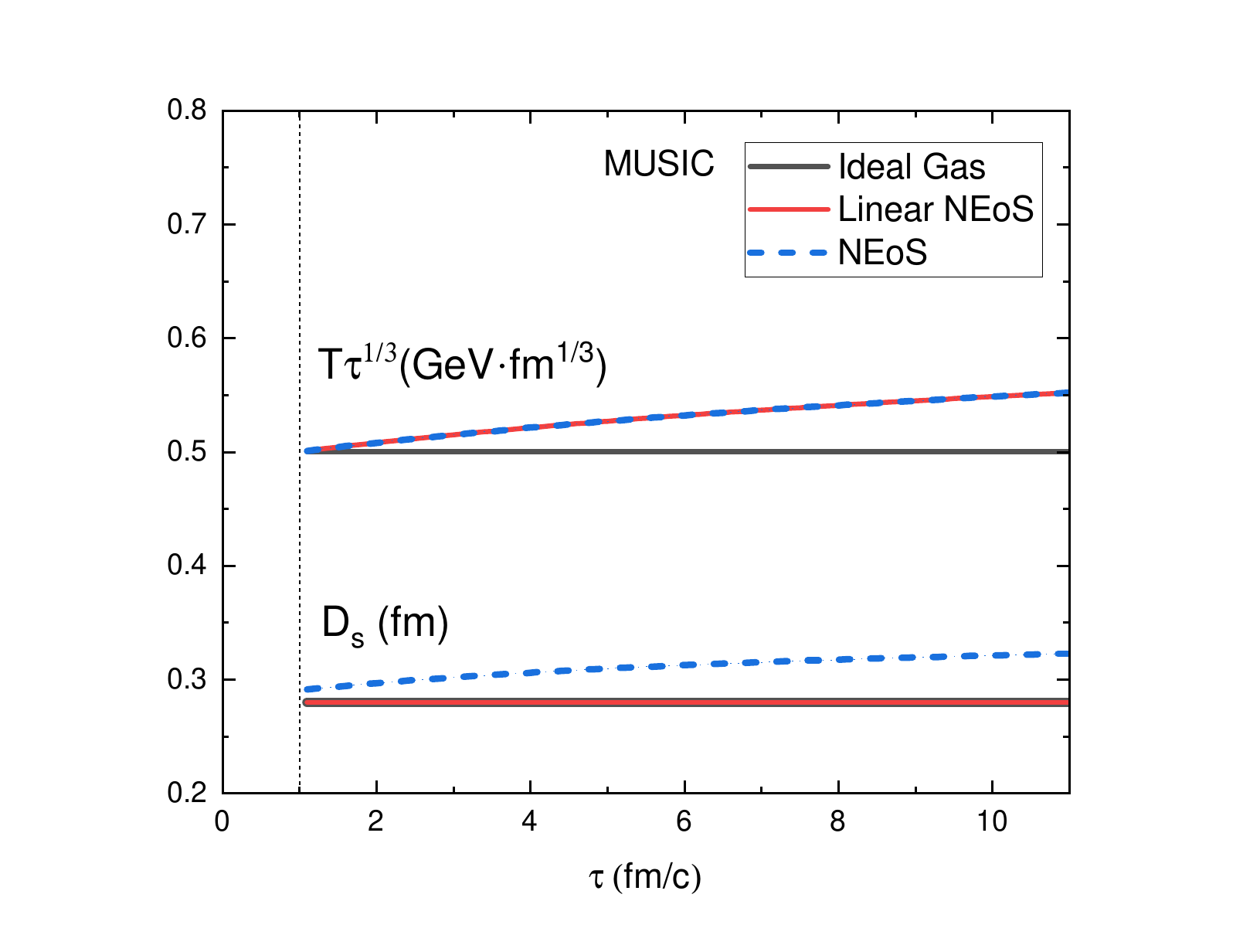}}
\subfigure{\centering
    \includegraphics[width=0.47\textwidth, trim= 0 0 0 0, clip]{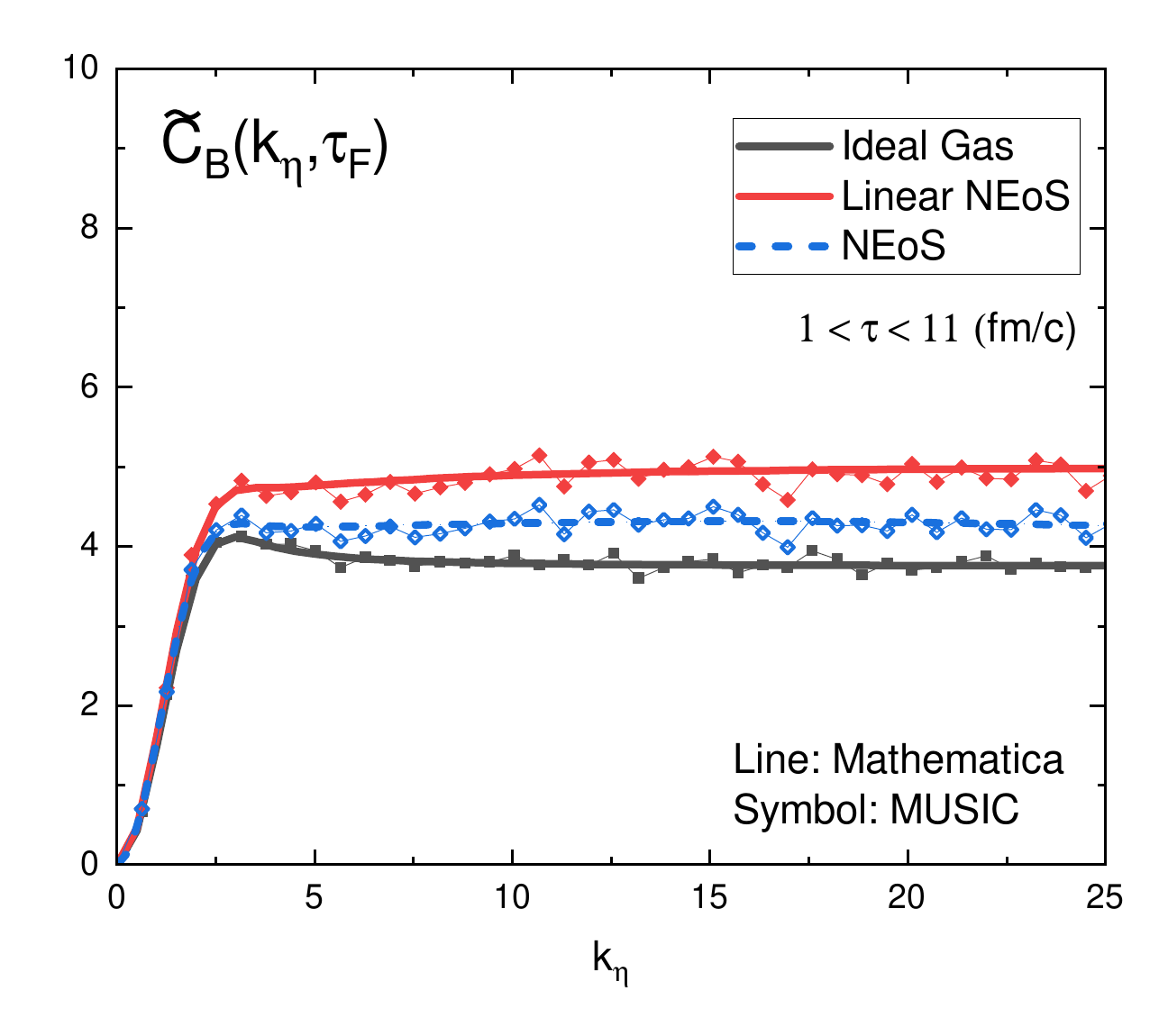}}\vspace{-4mm}
    \caption{Left: Evolution of temperature and diffusion constant for different equation of states. 
    Right: The corresponding two-point correlation at final time $\tau_F = 11~$fm/$c$. ``Linear NEoS" implements the temperature $T(e, n_B)$ from NEoS while $\mu_B(e, n_B)$ is same as for an ideal gas.  The transport coefficients $\kappa_B$ and $\tau_q$ are the same as in Fig.~\ref{fig:2pc_music}.}
    \label{fig:2pc_eos}
\end{figure*}

We would like to emphasize that the linearized approach allow multiple fluctuating baryon-transport events to run parallel based on a single background hydrodynamical event.
This dramatically reduces the computation time for stochastic hydrodynamics with sufficient resolution: To generate the data of Fig.~\ref{fig:2pc_music}, only 4 hours were needed to simulate the 10,000 events on a laptop. 

As a next step, we study the dependence of the two-point correlation on the choice of equation of state, baryon-number diffusion constant $\kappa_B$ and relaxation time $\tau_q$, respectively. 
Figure~\ref{fig:2pc_eos} compares the two-point correlation with different EoS. The left panel shows the time dependence of temperature and diffusion constant $D_s$. For the ideal gas EoS, both $T\tau^{1/3}$ and $D_s$ are constant, as follows from the assumptions in the 1+1D analytical calculations. 
However, such assumptions are too crude for a realistic hydrodynamic simulation of a heavy-ion collision, in particular when considering the softening of the speed of sound near the crossover. 
For a better description, we also consider the crossover equation of state NEoS-BQS \cite{Monnai:2021kgu}, which is constructed based on lattice QCD and a hadron resonance gas at zero chemical potential, and extended to the finite chemical potential region with a Taylor expansion. As the dashed lines show, the NEoS yields a moderate temperature evolution because of the smaller speed of sound. Also $D_s$ is no longer constant as the scaling $\chi_B \propto T^2$ is not satisfied. 
For comparison, we further consider the ``linear NEoS" that has the same temperature $T(e,n_B)$ as in NEoS-BQS, yet modifies the susceptibility using $\chi_B \propto T^2$ to reproduce the constant $D_s$ of the ideal gas EoS case. 

\begin{figure*}
\subfigure{\centering
    \includegraphics[width=0.49\textwidth, trim= 70 10 50 62, clip]{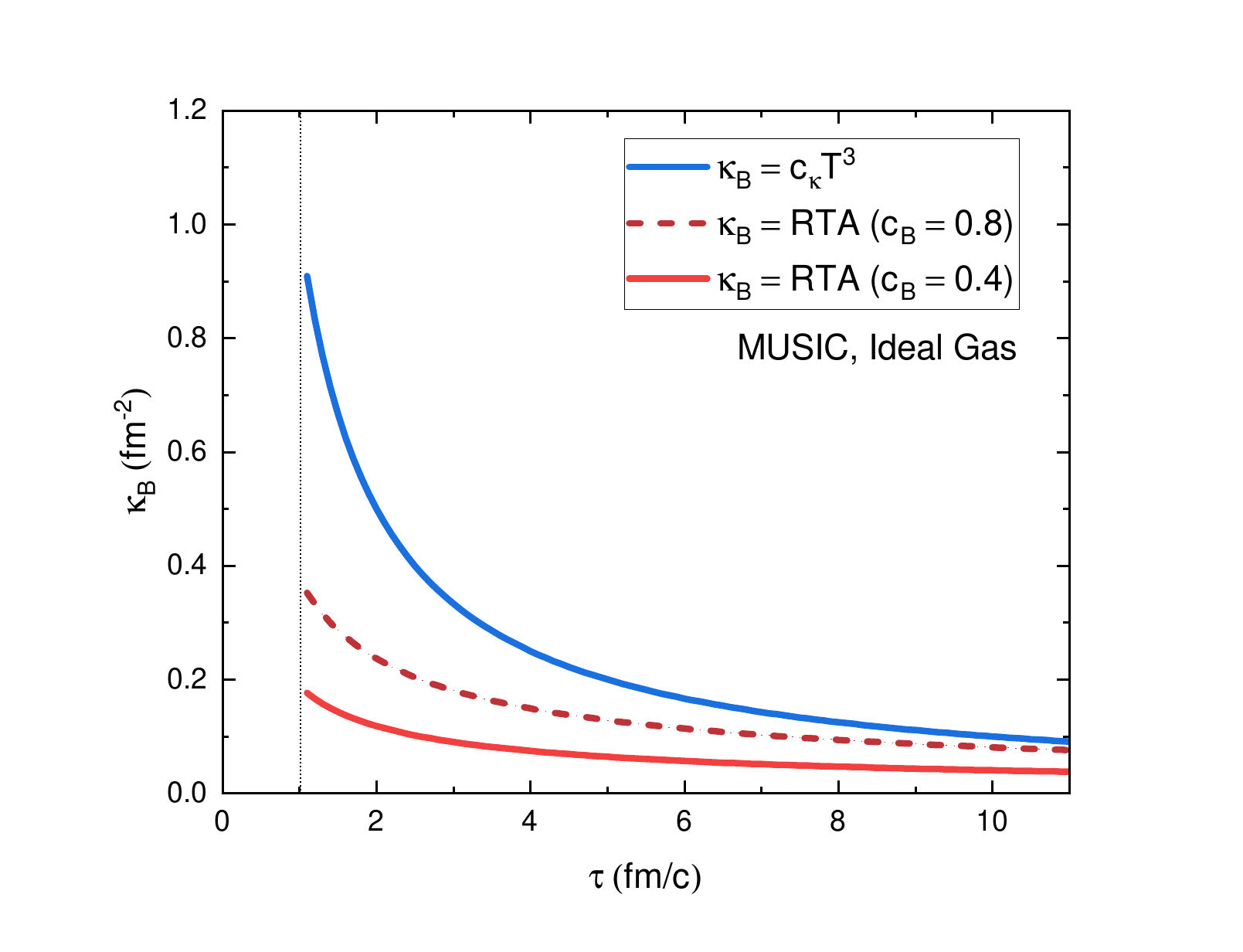}}
\subfigure{\centering
    \includegraphics[width=0.47\textwidth, trim= 0 0 0 0, clip]{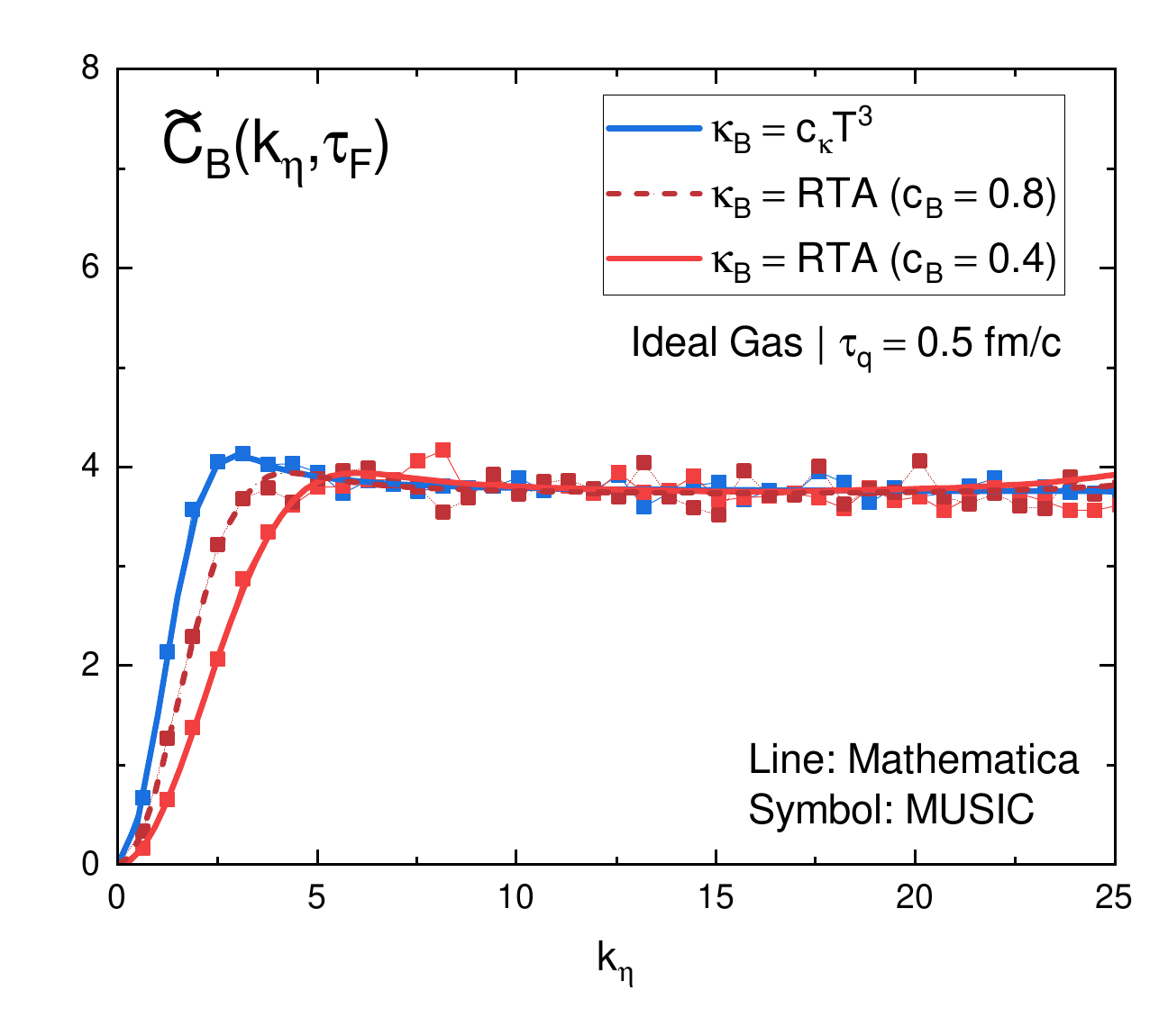}}\vspace{-4mm}
    \caption{Left: Time evolution of the baryon diffusion constant $\kappa_B$. 
    Right: The corresponding two-point correlation with final time $\tau_F = 11~$fm/$c$. We use the ideal gas EoS and the constant relaxation time $\tau_q = 0.5~$fm/$c$. 
    }
    \label{fig:2pc_kappa}
\end{figure*}

\begin{figure}
    \centering
    \includegraphics[width=0.48\textwidth, trim= 0 0 0 0, clip]{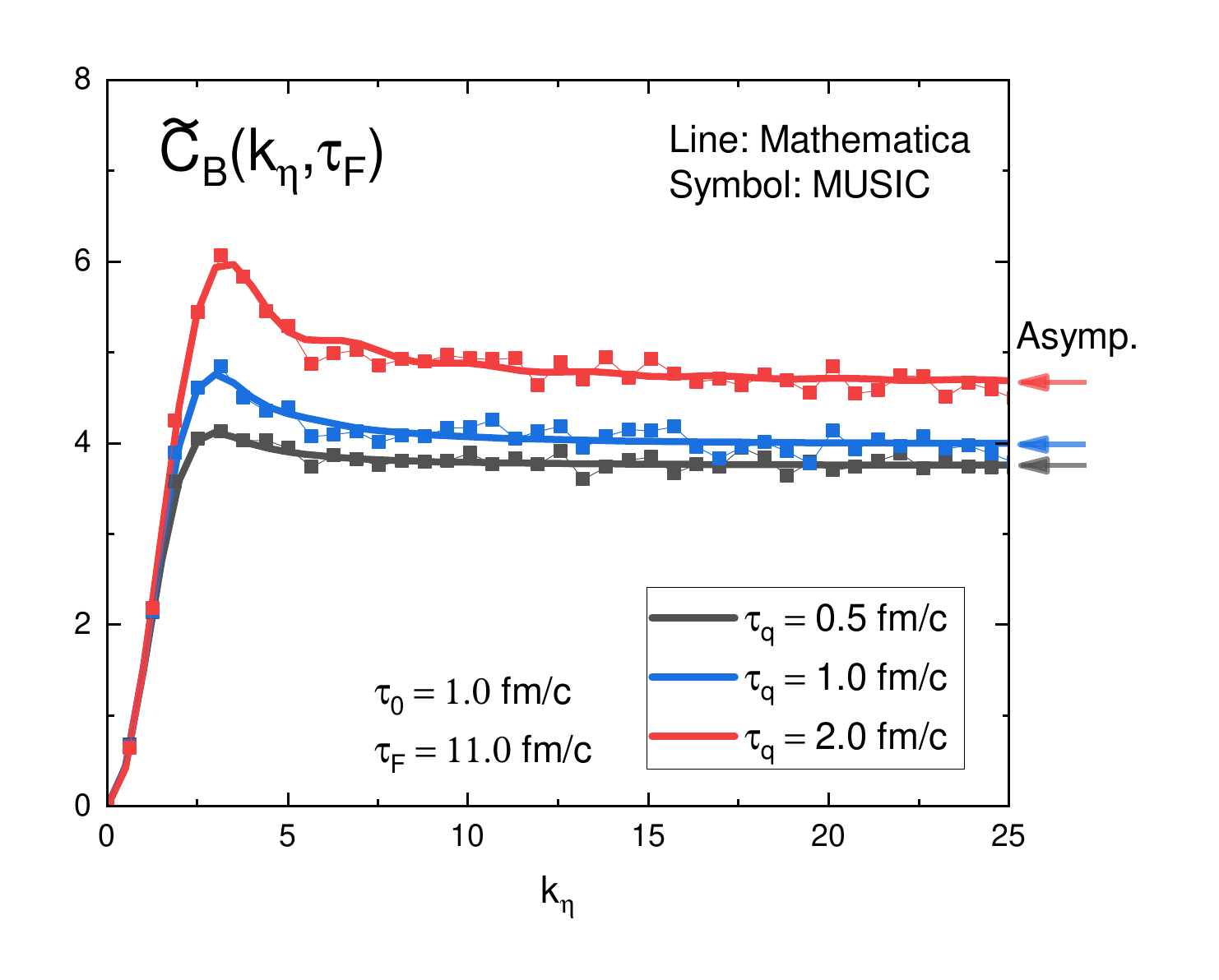}\vspace{-4mm}
    \caption{Dependence on the relaxation time $\tau_q$ of the two-point correlation. The symbols denote the MUSIC results and the lines denote the Mathematica calculation with different $\tau_q$, respectively.
    The arrows on the right show the asymptotic values deduced from the approximate solution~\eqref{eq:ei} valid at large $k_\eta$.}
    \label{fig:2pc_tauq}
\end{figure}
\begin{figure*}
\subfigure{\centering
    \includegraphics[width=0.48\textwidth, trim= 0 0 0 0, clip]{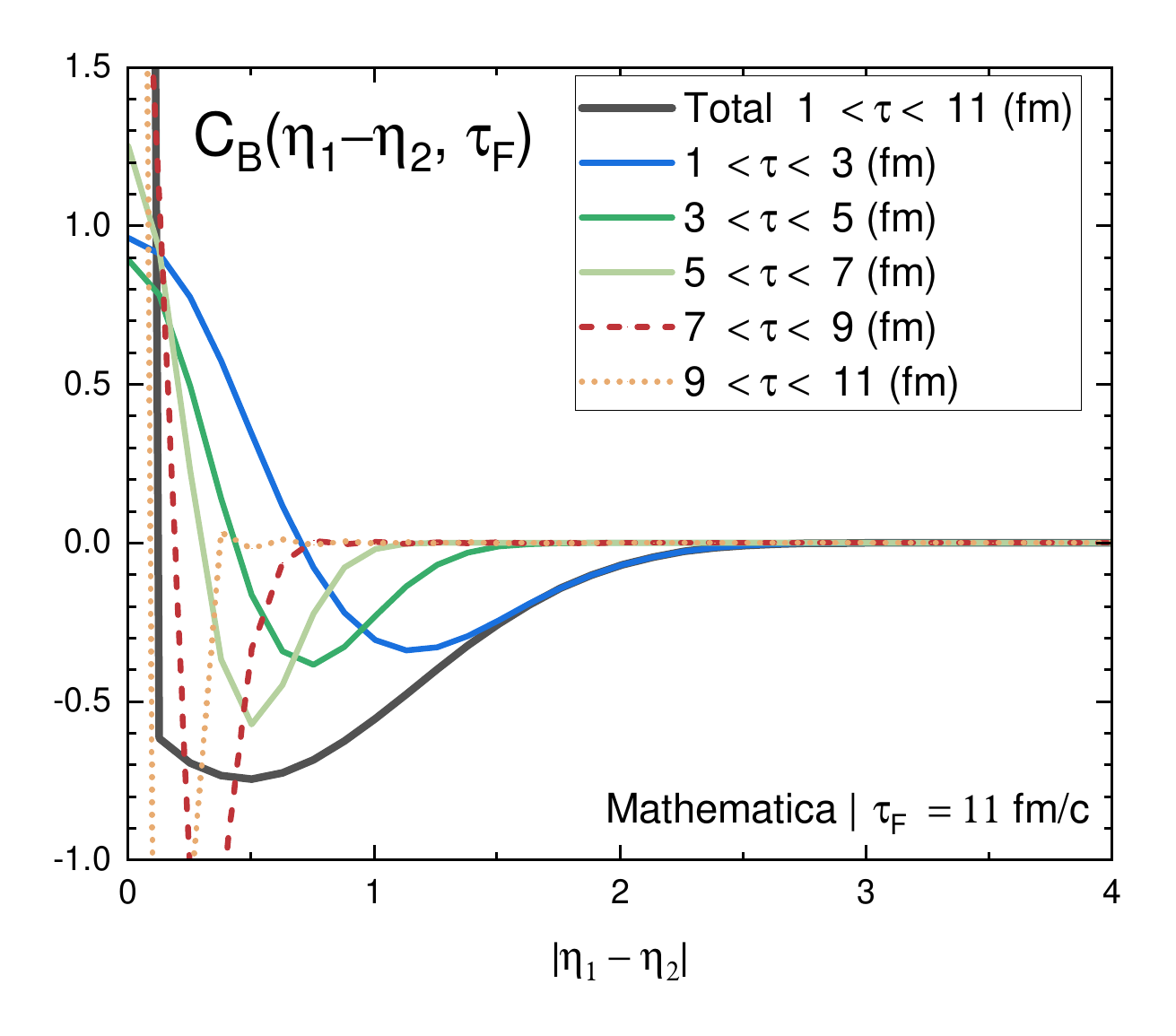}}
\subfigure{\centering
    \includegraphics[width=0.48\textwidth, trim= 70 0 70 55, clip]{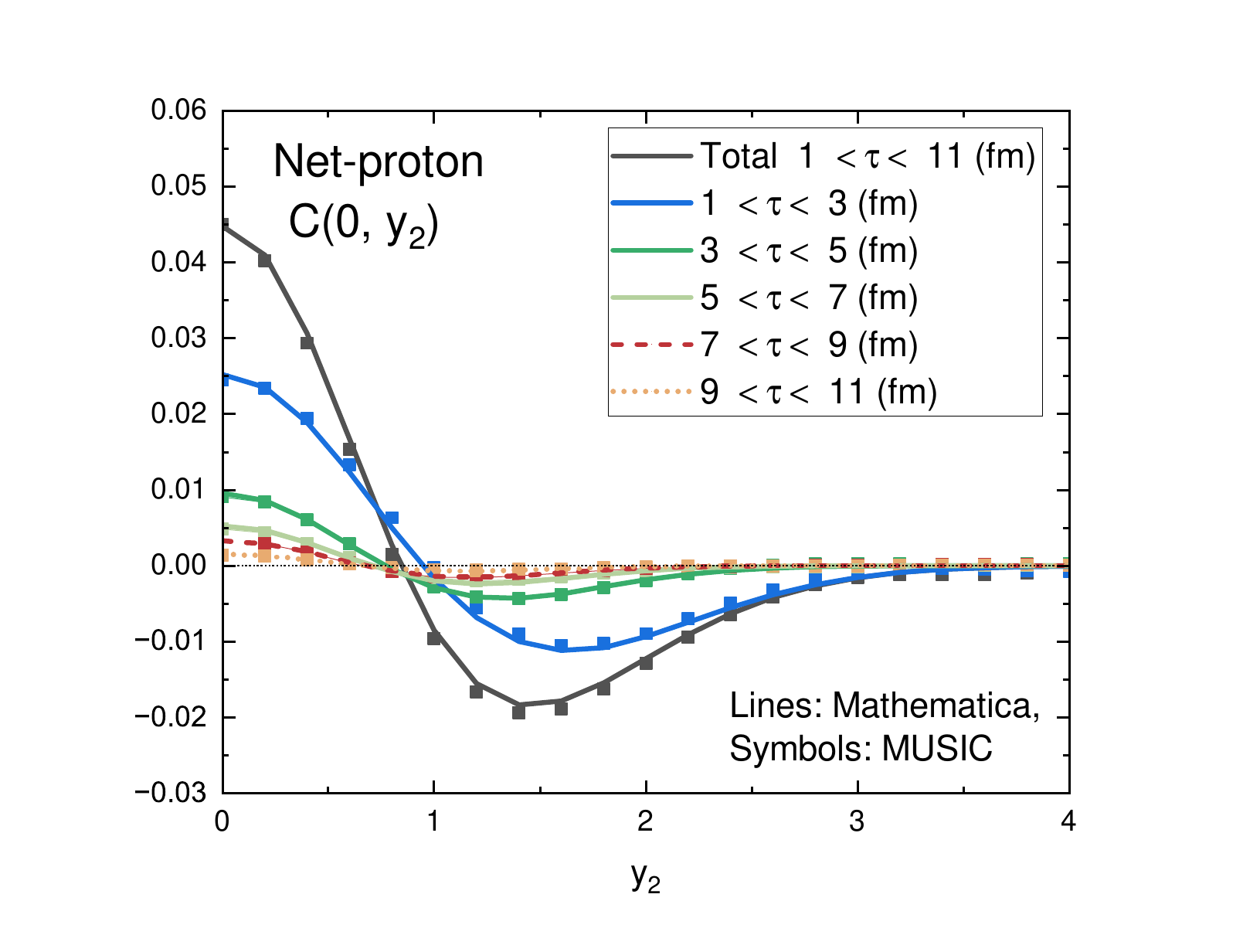}}
\vspace{-4mm}
    \caption{Longitudinal two-point correlation $C_B(\eta_1 - \eta_2, \tau_F)$ (left panel) and net-proton correlation $C(0, y_2)$ at freeze-out (right panel). The colored lines show the result with same final time $\tau_F = 11$~fm/$c$ but with different initial times, as in Fig.~\ref{fig:2pc_music}. 
    The effect of late-time fluctuations are plotted with dashed lines because of the numerical instabilities from the Dirac delta function.}
    \label{fig:2p_music}
\end{figure*}

As a result of these different behaviors, the right panel of Fig.~\ref{fig:2pc_eos} shows the corresponding two-point correlation $\tilde{C}_B(k_\eta, \tau_F)$ with ideal gas EoS, NEoS and linear NEoS, respectively. 
In the NEoS and linear NEoS cases, we extracted the $\kappa_B(\tau)$ evolution from the MUSIC profiles and used it as input for the Mathematica calculation.\footnote{Note that the NEoS limits the maximum chemical potential, $|\mu_B| \le 600~$MeV. Due to the large baryon charge fluctuations, the local chemical potential may be outside of the validity of the ordinary NEoS. Within this linear formalism, we re-scale the input noise for ordinary NEoS with a factor ${\rm d}\tau {\rm d}\eta$ and multiply by $1/({\rm d}\tau {\rm d}\eta)$ when calculating the two-point correlations. }  
The MUSIC results nicely reproduce the semi-analytical two-point correlation for all three equations of state. In particular the NEoS provides a more important late stage effect due to the softer EoS and slower temperature decrease, which is well captured by the MUSIC simulation.
Different equations of state show similar behavior in the low momentum part, where the evolution is dominated by the hydrodynamic mode and can be characterized by the baryon diffusion constant $\kappa_B$. 
At large $k_\eta$, the result demonstrates a clear ordering when changing the EoS. As the relaxation time $\tau_q = 0.5~$fm/$c$ is relatively small, this ordering can be understood from the $T \chi_B \tau_F$ ordering in  the Navier--Stokes limit~(\ref{eq:1d-corr-ns}). 

In Fig.~\ref{fig:2pc_kappa} we show the dependence of the two-point correlation on the transport coefficient $\kappa_B$. In the 1+1D reference calculation, we assume the scaling $\kappa_B \propto T^3$ which yields a constant $D_s$. 
We also used the more realistic temperature-dependence of $\kappa_B$ derived from the relaxation time approximation (RTA) \cite{Denicol:2018wdp}
\begin{equation}
\kappa_B = \frac{c_B}{T} n_B
    \bigg[ \frac{1}{3} \coth\big( \frac{\mu_B}{T} \big) - \frac{n_B T}{e + p} \bigg], 
\end{equation}
where the coefficient $c_B$ is a free parameter that controls the magnitude of the fluctuation and diffusion. 
The left panel of Fig.~\ref{fig:2pc_kappa} shows the time evolution of $\kappa_B$ with these different settings. The RTA $\kappa_B$ is relatively small, in particular at early times. 
As a consequence, the different choices of $\kappa_B$ lead to obvious differences in the two-point correlation in the low-$k_\eta$ region (right panel). 
In contrast, the asymptotic $\tilde{C}_B(k_\eta, \tau_F)$ at large $k_\eta$ is independent of $\kappa_B$, which can be found from the asymptotic solution~(\ref{eq:ei}). 

Figure~\ref{fig:2pc_tauq} shows the sensitivity of the two-point correlation on the relaxation time $\tau_q$. As already seen in the analytical results presented in Fig.~\ref{fig:1+1-2pc}, the two-point correlation has a roughly universal behavior at small $k_\eta$ and a $\tau_q$-dependent ordering at large $k_\eta$. 
The MUSIC calculations reproduce this ordering with different $\tau_q$, where the increasing relaxation time leads to a longer population of the non-hydrodynamic modes, and thus gradually deviates from the Navier--Stokes limit, showing a non-monotonic behavior with $k_\eta$. 

\section{Particle correlation after freeze-out}\label{sec:frz}

In Sec.~\ref{sec:2pc}, we calculated the two-point correlation of baryon charge in our extension of MUSIC with stochastic baryon-number transport, and compared the results with (semi-)analytical solutions. In order to relate the findings to experimental observables, such macroscopic hydrodynamic fields need to be converted into final hadron distributions.\footnote{For some analytical work on the freeze out of fluctuations, in the vicinity of the QCD critical point, see Ref.~\cite{Kapusta:2012zb}, the Hydro+ approach of Ref.~\cite{Pradeep:2022mkf}, and the blast-wave model in Ref.~\cite{Aasen:2022cid}.} 
In this section, we will discuss the correlation of primary hadrons arising from baryon-charge fluctuations. 

The traditional approach of this conversion is based on the Cooper--Frye freeze-out prescription at a switching hypersurface $\Sigma$~\cite{Cooper:1974mv}:
\begin{equation}\label{eq:cooperfrye}
    E\frac{{\rm d}N_i}{{\rm d}^3 p} = \frac{{\rm d}N_i}{p_T dp_T d\phi dy} = 
    \frac{g_i}{(2\pi)^3} \!\int_\Sigma p^\mu {\rm d}\sigma_\mu f_i(x,p),
\end{equation}
where the Lorentz-invariant particle distribution is given by the integration of the distribution function $f_i(x,p)$ over the hypersurface, with $g_i$ the degeneracy factor of the particle type $i$ and ${\rm d}\sigma_\mu$ the hypersurface element. 
Since the particles come from an almost equilibrated fluid, we use the Fermi--Dirac distribution function $f(x,p) = 1/[e^{ (p \cdot u \mp q_B\mu_B)/T } - 1]$, where $q_B$ denotes the baryon charge, while $-$ and $+$ are for baryons and anti-baryons, respectively. 
At linear order, the switching surface is determined by a fixed temperature $T = T_F$ of the background part. 
In a homogeneous system with Bjorken expansion, this switching condition simplifies to a fixed final time $\tau = \tau_F$. 
Note that in stochastic hydrodynamics, the local baryon chemical potential distribution $\mu_B(x)$ fluctuates event-by-event, and we employ its interpolated value at the center of each freeze-out surface patch to calculate the event-averaged observables. To compare with the 1+1D analytical result, we focus on the longitudinal particle distribution that characterizes the longitudinal evolution:
\begin{equation}
    \frac{{\rm d}N_i}{{\rm d}y} = \frac{g_i}{(2\pi)^3} 
    \!\int\! p_T\, {\rm d}p_T {\rm d}\phi \int\! p^\mu {\rm d}\sigma_\mu f_i(x,p).
\end{equation}
With white noise input, we expect the event-averaged effect on the longitudinal spectra to vanish while the variance, and more generally the two-particle correlation in momentum-space rapidity, is non-zero due to the finite two-point correlation in spacetime rapidity. 
Such a calculation is straightforward in a numerical hydrodynamic framework, while we need to further simplify this scheme in the analytical solution used for comparison. 
At the leading order, the two-particle correlation only receives contributions from the fluctuations of the distribution function. 
We consider the fluctuation of baryon number density as a perturbation 
\begin{equation}\label{eq:delta_dN/dy}
    \delta\!\left( \frac{{\rm d}N_i}{{\rm d}y} \right ) =
    \frac{g_i}{(2\pi)^3} 
    \!\int\! p_T\, {\rm d}p_T\,{\rm d}\phi \!\int\! p^\mu {\rm d}\sigma_\mu\,\frac{\partial f(x,p)}{\partial (\mu_B/ T)} \frac{\delta n_B}{ \chi_{B} T_F},
\end{equation}
where we used the relation $\chi_B = \partial n_B / \partial \mu_B$.  The corresponding two-particle longitudinal correlation is defined as:
\begin{equation}
C(y_1, y_2) \equiv \left \langle\delta\!\left( \frac{dN_1}{dy_1} \right ) \delta\!\left ( \frac{dN_2}{dy_2} \right ) \right \rangle \Big / ( dN/dy ).
\end{equation}
Inserting Eq.~\eqref{eq:delta_dN/dy}, this yields after some calculation
\begin{align}
C(y_1, y_2) = &\bigg[\frac{g_i}{(2 \pi)^3 \chi_f \tau_f T_f} \bigg]^2 
  \!\int\!{\rm d}^2{\bf p}_{T,1} \int\!{\rm d}^2{\bf p}_{T,2} \cr
  &\int\!p^{\mu\,}_1{\rm d}\sigma_{\mu,1} \!\int\!p^{\mu\,}_2{\rm d}\sigma_{\mu,2}\;  
  f'_1(x_1, p_1) f'_2(x_2, p_2) \cr
  &\times \frac{C_B(\eta_1 - \eta_2, \tau_F)}{\frac{1}{S_\bot}(dN/dy)}
\end{align}
where $f' = \partial f / \partial (\mu_B/T)$ denotes the derivative of the Fermi--Dirac distribution, $S_\bot$ is the transverse area of the system, and the two-point correlation $C_B$ in coordinate space is given by the inverse Fourier transform of $\tilde{C}_B(k_\eta, \tau_F)$:
\begin{equation}
\label{eq:1d-corrx}
C_B(\eta_1 - \eta_2, \tau_F)
       = \int\!\frac{{\rm d} k_\eta}{2\pi}\,  
       e^{{\rm i}k_{\eta}(\eta_1-\eta_2)}
       \tilde{C}_B(k_\eta, \tau_F).
\end{equation}

\begin{figure}
    \centering
    \includegraphics[width=0.48\textwidth, trim= 50 0 50 0, clip]{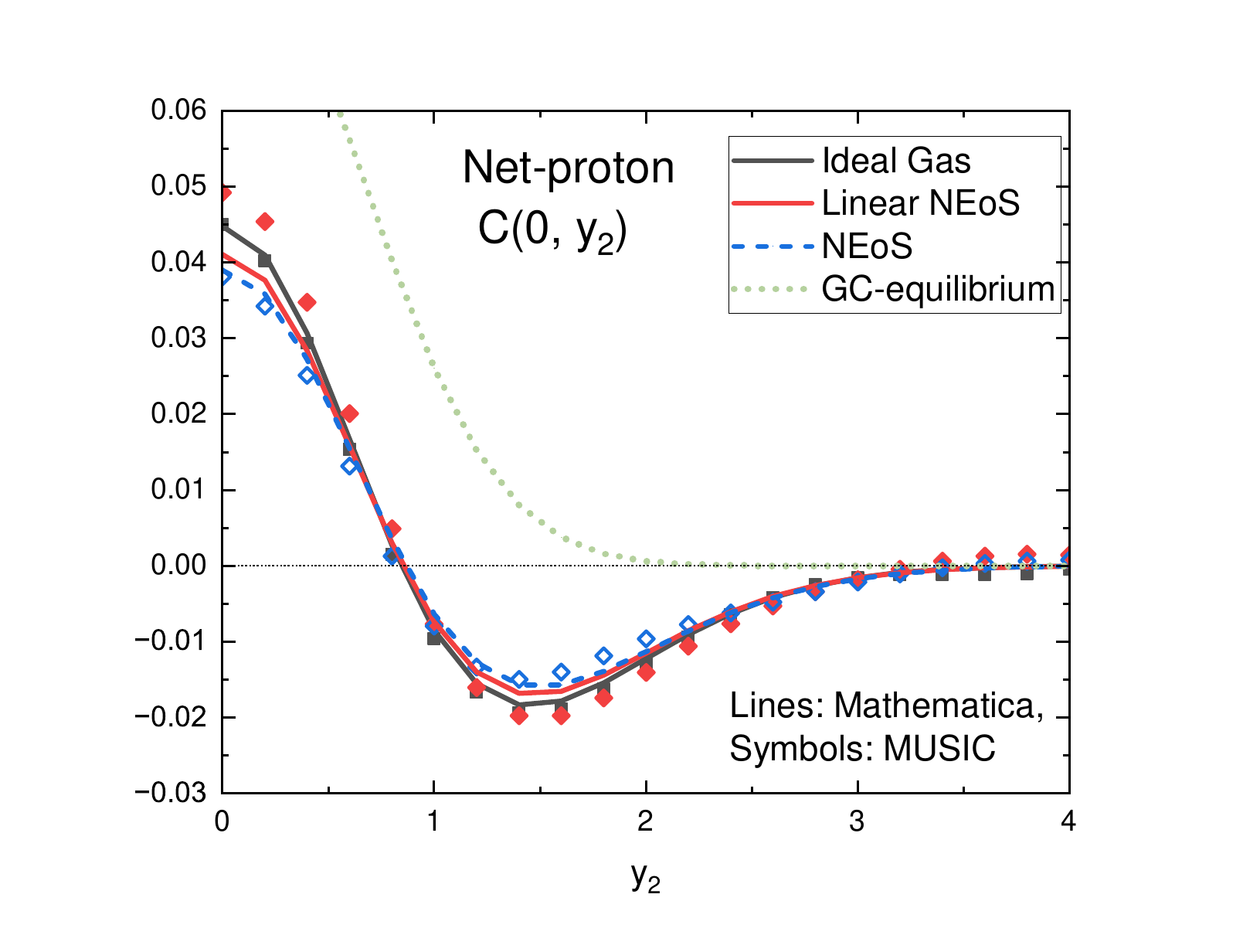}\vspace{-4mm}
    \caption{Dependence of the net-proton correlation at the freeze out on the equation of state. The dotted line shows the result for equilibrium charge fluctuations in a grand canonical ensemble.}
    \label{fig:2p_eos}
\end{figure}
\begin{figure}
    \centering
    \includegraphics[width=0.48\textwidth, trim= 50 0 50 0, clip]{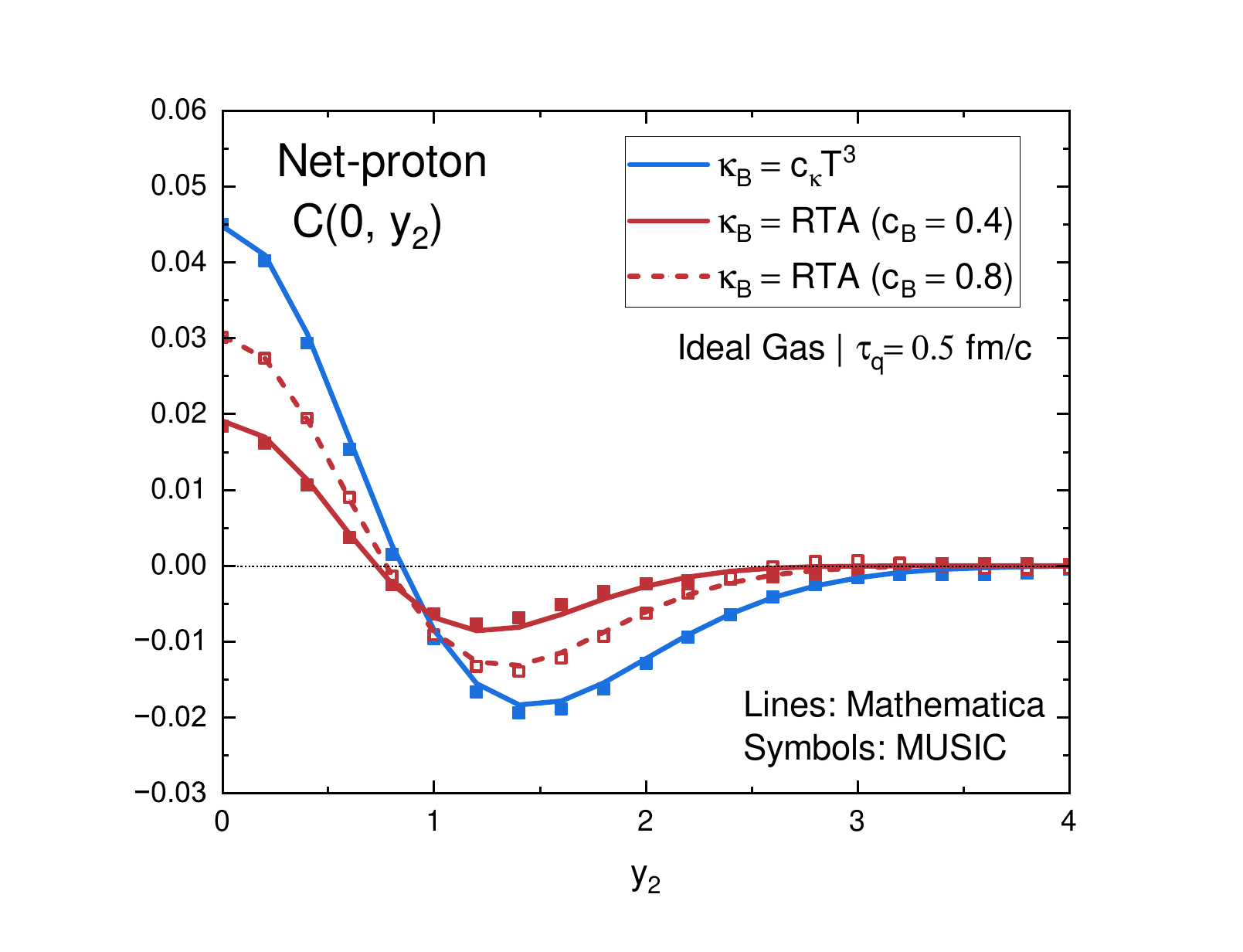}\vspace{-4mm}
    \caption{Dependence on the charge diffusion coefficient $\kappa_B$ of the net-proton correlation at freeze out.}
    \label{fig:2p_kappa}
\end{figure}
\begin{figure}
    \centering
    \includegraphics[width=0.48\textwidth, trim= 50 0 50 0, clip]{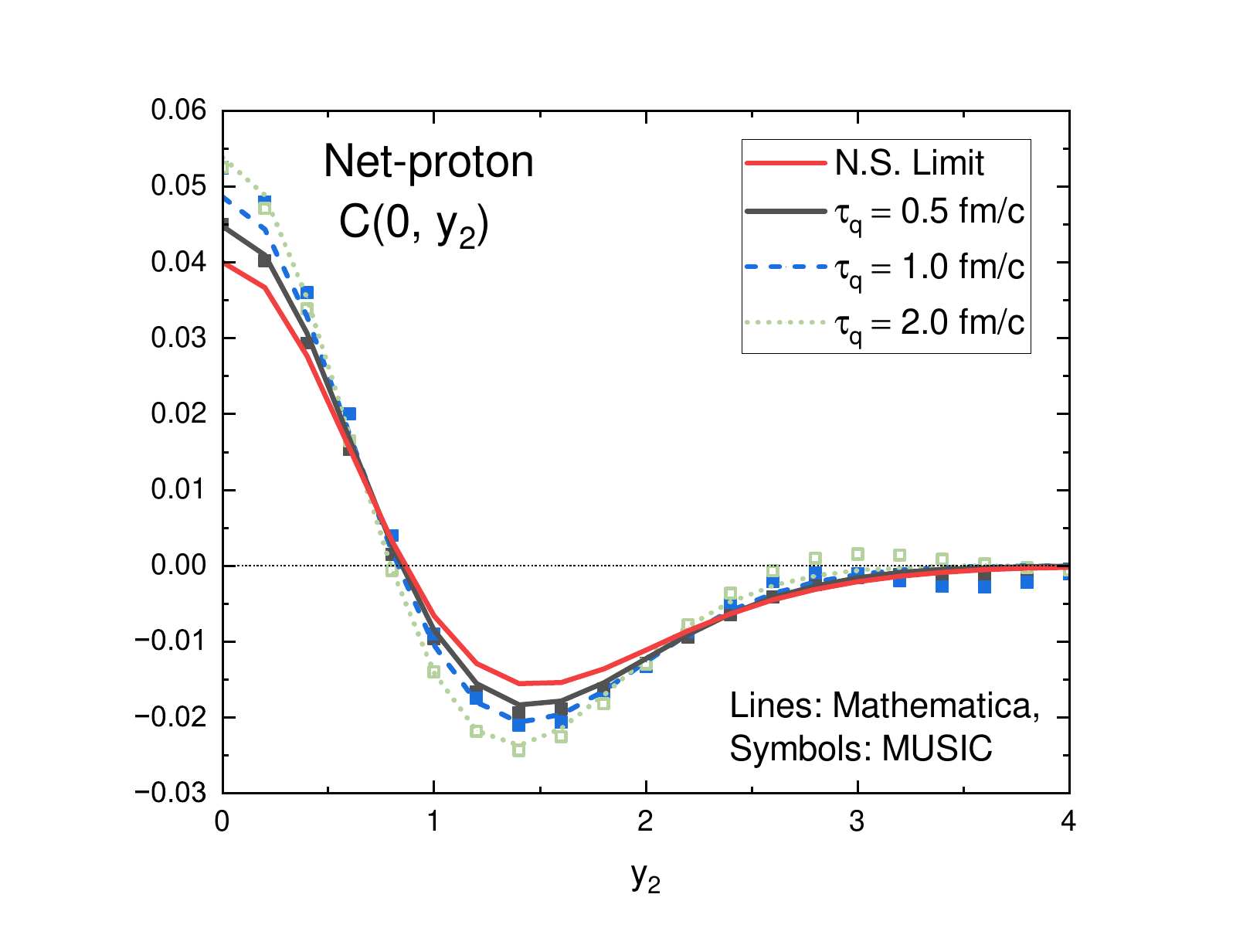}\vspace{-4mm}
    \caption{Dependence of the net-proton correlation at freeze out on the diffusion relaxation time.}
    \label{fig:2p_tauq}
\end{figure}

Figure \ref{fig:2p_music} shows the coordinate space correlation $C_B(\eta_1 - \eta_2, \tau_F)$ (left panel) and the corresponding net-proton correlation $C(0, y_2)$ (right panel), including the separate contributions of fluctuations originating in different time intervals, as done in Fig.~\ref{fig:2pc_music}. 
We find that the total correlation $C_B(\eta_1 - \eta_2, \tau_F)$ shows a striking peak at $|\eta_1 - \eta_2| = 0$. In the limit $\tau_q \rightarrow 0$, this peak would be a Dirac-delta in the Navier--Stokes theory, which was referred to as self-correlation in Refs.~\cite{Ling:2013ksb,Pratt:2012dz,De:2020yyx}. 
For $|\eta_1 - \eta_2| > 0$, $C_B(\eta_1 - \eta_2, \tau_F)$ is negative due to overall charge conservation. 
The influence of the difference in the time at which noise emerges leads to an ordering of the correlation scale, with the earlier fluctuations spreading at longer distances, corresponding to the lower momentum region in Fig.~\ref{fig:2pc_music}. 
Such an ordering of two-point correlations can also be found after Cooper--Frye particlization in the right panel, where MUSIC simulations nicely match the analytical results. 
However, the ordering according to creation time (or equivalently $k_\eta$) is smeared by the Cooper--Frye freeze out such that in particular, the late stage noise that is significant at large $k_\eta$ only contributes weakly to the two-particle correlations. Nevertheless, the Cooper--Frye prescription only provides an average description of the freeze-out dynamics, and a more comprehensive description of particlization and freeze out is required in the future to evaluate the impact of the fluctuations.

Figure \ref{fig:2p_eos} compares the two-particle correlation $C(0, y_2)$ computed with different equations of state. Since the EoS mostly affects the large-momentum region (Fig.~\ref{fig:2pc_eos}), these differences disappear after the Cooper--Frye freeze out and the various EoS result in a roughly universal behavior. 
We can understand this result using the Navier--Stokes form~(\ref{eq:1d-corr-ns}), in which the constant term $\chi_F T_F \tau_F$ corresponds to the charge fluctuations in thermal equilibrium, while the exponential term provides a negative contribution as a result of charge conservation. 
For comparison, we also show as dotted line the equilibrium charge fluctuations in a grand-canonical ensemble, which are largely suppressed by the non-equilibrium dynamical evolution. 

In contrast, the different two-point correlations $\tilde{C}_B(k_\eta,\tau_F)$ at low $k_\eta$ observed in Fig.~\ref{fig:2pc_kappa} for various choices of $\kappa_B$  translate into a significant difference in the longitudinal two-particle correlation shown in Fig.~\ref{fig:2p_kappa}, and thus survive after freeze out. 
This implies that fluctuating observables can possibly be used to probe the transport property of QCD matter and constrain the charge diffusion coefficient.  

Eventually, Fig.~\ref{fig:2p_tauq} displays the dependence of $C(0, y_2)$ on the relaxation time $\tau_q$. With increasing $\tau_q$, the two-particle correlation increases at small rapidity difference, which can be attributed to the ordering at large $k_\eta$ observed in Fig.~\ref{fig:2pc_kappa}, and we find a nice agreement between the MUSIC + Cooper--Frye calculations and the analytical ones. 
However, the current particlization procedure mostly reflects the structure at low $k_\eta$ that is dominated by the Navier--Stokes regime, and a more comprehensive study of the influence of relaxation times is needed in the future.

\section{Summary} \label{sec:sum}

In this work, we studied the fluctuations and transport of baryon charge in a linearized stochastic hydrodynamic approach. We extended the 3+1D hydrodynamic code MUSIC to include the fluctuations of baryon number and its diffusion on top of the ordinary energy-momentum evolution. This linearized approach provides reliable simulations and nicely describes the two-point correlations with various choices of equations of state and transport coefficients. In particular, it largely reduces the computational requirement of the fluctuating hydrodynamics and can be applied event-by-event in realistic 3+1D simulations of heavy-ion collisions in the future.  

The hydrodynamic calculations demonstrate how the final observables relate the noise created at different times in the evolution to correlations over different rapidity range. The long-range longitudinal correlations are mostly induced by the early-stage noise and are close to results from first-order Navier--Stokes diffusion. In turn, the low-rapidity-separation part is dominated by the late stage effects, and is affected by the equation of state and the relaxation time, possibly providing an avenue to assess the latter, i.e.\ an otherwise poorly constrained second-order coefficient.

The linearized stochastic hydrodynamics also nicely describes the two-particle correlation after Cooper--Frye freeze out. It shows particular sensitivity on the charge diffusion constant $\kappa_B$, and may be applied to constrain this important transport coefficient from experimental measurements. 
However, our current Cooper--Frye implementation is still insufficient to translate the correlations during freeze out, and a more comprehensive description is necessary. 
In the future, we expect to apply this linear stochastic hydrodynamics to realistic 3+1D heavy-ion collisions and to incorporate initial state fluctuation and freeze-out fluctuations, to see which observables still retain information from the thermal fluctuations in the hydrodynamic evolution.

\section*{Acknowledgments}

We thank Anton Andronic, Fei Gao, and Hendrik Roch for insightful discussions. 
This work is supported by the programme Netzwerke 2021, an initiative of the Ministry of Culture and Science of the State of Northrhine Westphalia (MKW NRW)
under the funding code NW21-024-A.
The sole responsibility for the content of this publication lies with the authors.
The authors acknowledge computing time provided by the Paderborn Center for Parallel Computing (PC2). 
\vspace{1cm}

\appendix
\section{Asymptotic Green's function}
\label{appendix:asymptotic}

In a 1+1D system with Bjorken expansion, the Green's function at large $k_\eta$ takes the asymptotic form~(\ref{eq:green-asymp}) with large $\nu$, which we recall here for convenience:
\begin{equation}\label{eq:asymp-green-appendix}
\tilde{G}_{\rm as}(k_\eta; s, s_0) = - 2\sqrt{ s s_0}\, \frac{e^{ s_0 - s}}{\nu} \sin\bigg(\nu\ln\frac{s}{s_0}\bigg). 
\end{equation}

\begin{figure}
    \centering
    \includegraphics[width=0.48\textwidth, trim= 50 0 50 0, clip]{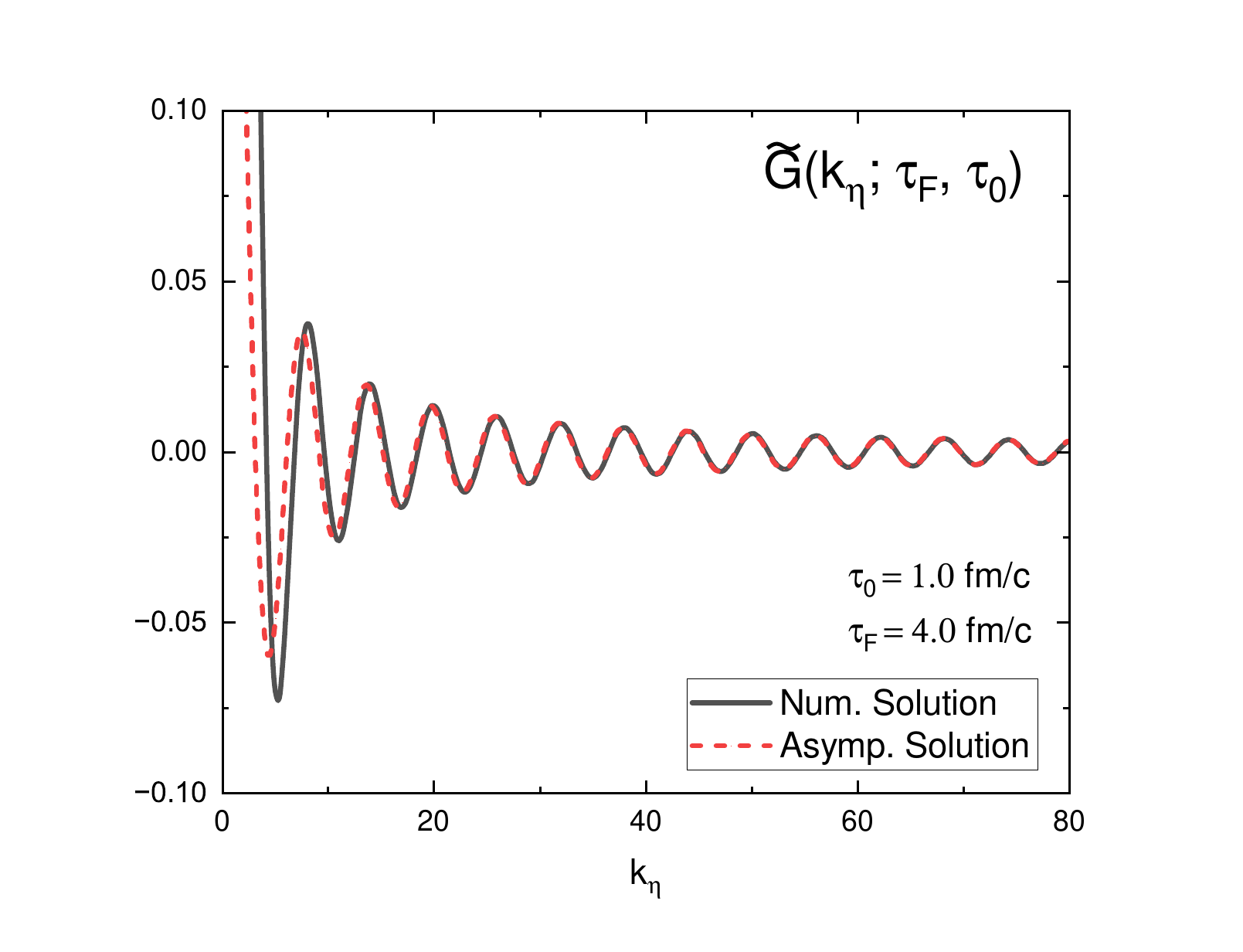}
    \caption{The Green's function in 1+1D system. The solid and dashed lines show the analytical result Eq.~(\ref{eq:1d_gf}) and asymptotic result Eq.~(\ref{eq:green-asymp}), respectively. }
    \label{fig:asymp-greenf}
\end{figure}

Figure \ref{fig:asymp-greenf} compares the exact Green's function and its asymptotic form. It shows the good consistency of Eq.~\eqref{eq:asymp-green-appendix} with the exact Green's function at large $nu$, which corresponds to large $k_\eta$ or relatively small $\tau_q$. 
Inserting $\tilde{G}_{\rm as}$ in Eq.~(\ref{eq:1d-cor}) for the two-point correlation at freeze out, and assuming the simplified time-dependence $\kappa_B(\tau) = \kappa_B(\tau_0) \tau_0/\tau$ for the diffusion constant, as done in the calculation of $\tilde{C}_B$ in Fig.~\ref{fig:1+1-2pc}, the integral can be performed analytically:
\begin{widetext}
\begin{equation}
\begin{aligned}\label{eq:ei}
\tilde{C}_{B}(k_\eta, \tau_f)
    &= \int^{\tau_f}_{\tau_0} \frac{{\rm d}\tau}{\tau^2}  k_\eta^2 
    \tilde{G}_{\rm as}^2(k_\eta; \tau_f, \tau) 2 \kappa_B(\tau_0)\tau_0\\
    &= \frac{\kappa_B(\tau_0)\tau_0}{2 D_s}  e^{-x_F} x_F \Big[ 2 \text{Ei}(x_0) + (- x_F)^{2 {\rm i} \nu} \Gamma(-2 {\rm i} \nu, -x_0) + (- x_F)^{-2 {\rm i} \nu} \Gamma(2 {\rm i} \nu, -x_0)   \Big],
\end{aligned}
\end{equation}
\end{widetext}
where Ei denotes the exponential integral.
The corresponding term provides a $k_\eta$-independent baseline for the large $k_\eta$ limit, while the two incomplete gamma functions represent a $k_\eta$-dependent oscillation about this baseline.

\begin{figure}
    \centering
    \includegraphics[width=0.48\textwidth, trim= 0 0 0 0, clip]{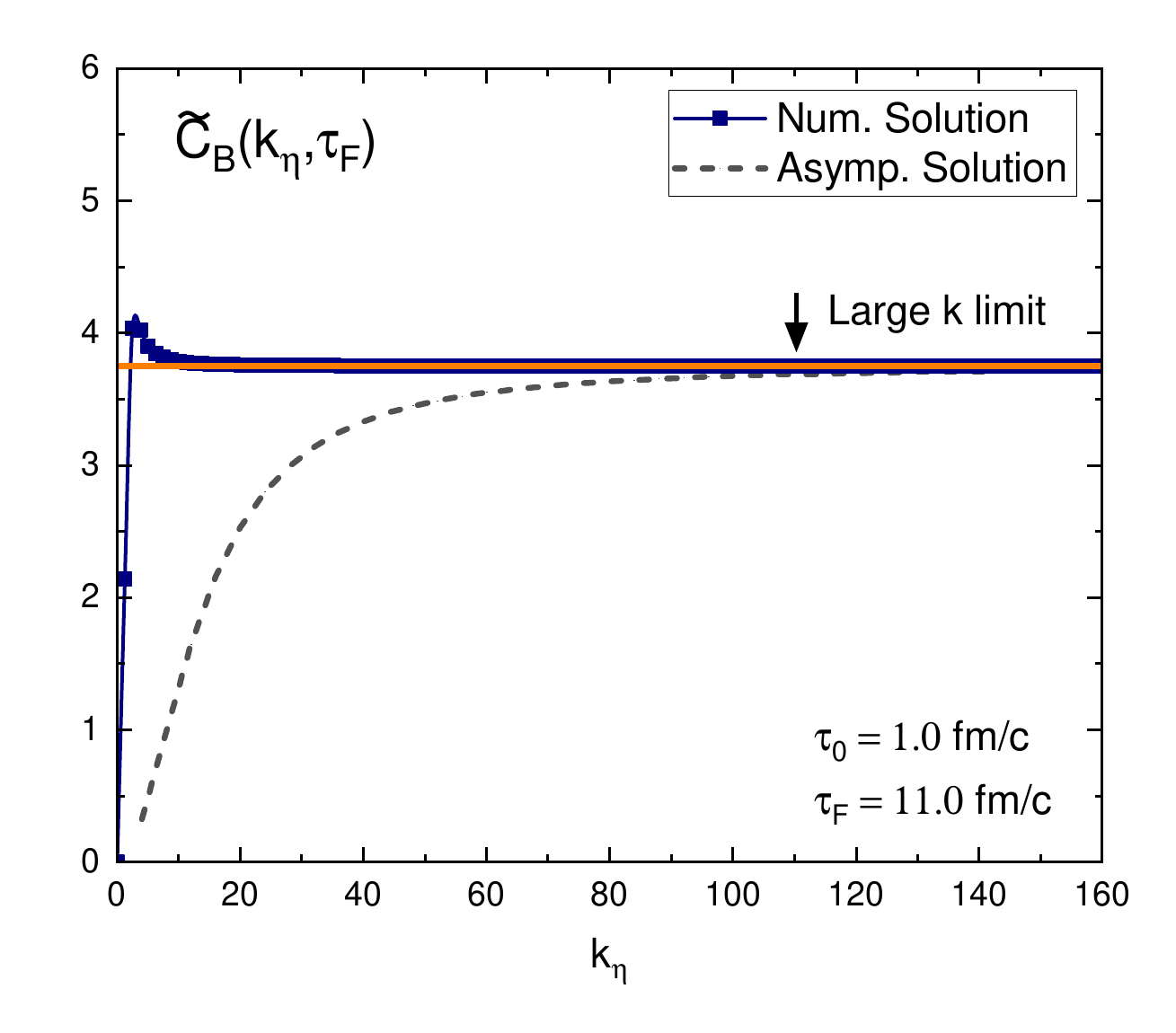}
    \caption{The two-point correlation from the analytical Green's function (symbols) and the asymptotic form (dashed lines). The orange line shows the large $k_\eta$ limit baseline from the aymptotic integration. }
    \label{fig:asymp-2pc}
\end{figure}

In Fig.~\ref{fig:asymp-2pc} we display the two-point correlation function from a numerical integration with the exact Green's function, together with the approximate form~\eqref{eq:ei}, towards which it converges at large $k_\eta$. 
We also show the baseline from the Ei function as the constant line labeled ``large k limit", which indeed correctly describes the high-$k_\eta$ behavior. 
As shown in Fig.~\ref{fig:1+1-2pc}, it also reproduces the asymptotic behavior in the Navier--Stokes case when $\tau_q$ is close to zero.


\bibliography{lfqgp}

\begin{thebibliography}{60}%
\makeatletter
\providecommand \@ifxundefined [1]{%
 \@ifx{#1\undefined}
}%
\providecommand \@ifnum [1]{%
 \ifnum #1\expandafter \@firstoftwo
 \else \expandafter \@secondoftwo
 \fi
}%
\providecommand \@ifx [1]{%
 \ifx #1\expandafter \@firstoftwo
 \else \expandafter \@secondoftwo
 \fi
}%
\providecommand \natexlab [1]{#1}%
\providecommand \enquote  [1]{``#1''}%
\providecommand \bibnamefont  [1]{#1}%
\providecommand \bibfnamefont [1]{#1}%
\providecommand \citenamefont [1]{#1}%
\providecommand \href@noop [0]{\@secondoftwo}%
\providecommand \href [0]{\begingroup \@sanitize@url \@href}%
\providecommand \@href[1]{\@@startlink{#1}\@@href}%
\providecommand \@@href[1]{\endgroup#1\@@endlink}%
\providecommand \@sanitize@url [0]{\catcode `\\12\catcode `\$12\catcode `\&12\catcode `\#12\catcode `\^12\catcode `\_12\catcode `\%12\relax}%
\providecommand \@@startlink[1]{}%
\providecommand \@@endlink[0]{}%
\providecommand \url  [0]{\begingroup\@sanitize@url \@url }%
\providecommand \@url [1]{\endgroup\@href {#1}{\urlprefix }}%
\providecommand \urlprefix  [0]{URL }%
\providecommand \Eprint [0]{\href }%
\providecommand \doibase [0]{http://dx.doi.org/}%
\providecommand \selectlanguage [0]{\@gobble}%
\providecommand \bibinfo  [0]{\@secondoftwo}%
\providecommand \bibfield  [0]{\@secondoftwo}%
\providecommand \translation [1]{[#1]}%
\providecommand \BibitemOpen [0]{}%
\providecommand \bibitemStop [0]{}%
\providecommand \bibitemNoStop [0]{.\EOS\space}%
\providecommand \EOS [0]{\spacefactor3000\relax}%
\providecommand \BibitemShut  [1]{\csname bibitem#1\endcsname}%
\let\auto@bib@innerbib\@empty
\bibitem [{\citenamefont {Aoki}\ \emph {et~al.}(2006)\citenamefont {Aoki}, \citenamefont {Endrodi}, \citenamefont {Fodor}, \citenamefont {Katz},\ and\ \citenamefont {Szabo}}]{Aoki:2006we}%
  \BibitemOpen
  \bibfield  {author} {\bibinfo {author} {\bibfnamefont {Y.}~\bibnamefont {Aoki}}, \bibinfo {author} {\bibfnamefont {G.}~\bibnamefont {Endrodi}}, \bibinfo {author} {\bibfnamefont {Z.}~\bibnamefont {Fodor}}, \bibinfo {author} {\bibfnamefont {S.~D.}\ \bibnamefont {Katz}}, \ and\ \bibinfo {author} {\bibfnamefont {K.~K.}\ \bibnamefont {Szabo}},\ }\href {\doibase 10.1038/nature05120} {\bibfield  {journal} {\bibinfo  {journal} {Nature}\ }\textbf {\bibinfo {volume} {443}},\ \bibinfo {pages} {675} (\bibinfo {year} {2006})},\ \Eprint {http://arxiv.org/abs/hep-lat/0611014} {arXiv:hep-lat/0611014} \BibitemShut {NoStop}%
\bibitem [{\citenamefont {Ding}\ \emph {et~al.}(2015)\citenamefont {Ding}, \citenamefont {Karsch},\ and\ \citenamefont {Mukherjee}}]{Ding:2015ona}%
  \BibitemOpen
  \bibfield  {author} {\bibinfo {author} {\bibfnamefont {H.-T.}\ \bibnamefont {Ding}}, \bibinfo {author} {\bibfnamefont {F.}~\bibnamefont {Karsch}}, \ and\ \bibinfo {author} {\bibfnamefont {S.}~\bibnamefont {Mukherjee}},\ }\href {\doibase 10.1142/S0218301315300076} {\bibfield  {journal} {\bibinfo  {journal} {Int. J. Mod. Phys. E}\ }\textbf {\bibinfo {volume} {24}},\ \bibinfo {pages} {1530007} (\bibinfo {year} {2015})},\ \Eprint {http://arxiv.org/abs/1504.05274} {arXiv:1504.05274 [hep-lat]} \BibitemShut {NoStop}%
\bibitem [{\citenamefont {Stephanov}\ \emph {et~al.}(1998)\citenamefont {Stephanov}, \citenamefont {Rajagopal},\ and\ \citenamefont {Shuryak}}]{Stephanov:1998dy}%
  \BibitemOpen
  \bibfield  {author} {\bibinfo {author} {\bibfnamefont {M.~A.}\ \bibnamefont {Stephanov}}, \bibinfo {author} {\bibfnamefont {K.}~\bibnamefont {Rajagopal}}, \ and\ \bibinfo {author} {\bibfnamefont {E.~V.}\ \bibnamefont {Shuryak}},\ }\href {\doibase 10.1103/PhysRevLett.81.4816} {\bibfield  {journal} {\bibinfo  {journal} {Phys. Rev. Lett.}\ }\textbf {\bibinfo {volume} {81}},\ \bibinfo {pages} {4816} (\bibinfo {year} {1998})},\ \Eprint {http://arxiv.org/abs/hep-ph/9806219} {arXiv:hep-ph/9806219} \BibitemShut {NoStop}%
\bibitem [{\citenamefont {Pandav}\ \emph {et~al.}(2022)\citenamefont {Pandav}, \citenamefont {Mallick},\ and\ \citenamefont {Mohanty}}]{Pandav:2022xxx}%
  \BibitemOpen
  \bibfield  {author} {\bibinfo {author} {\bibfnamefont {A.}~\bibnamefont {Pandav}}, \bibinfo {author} {\bibfnamefont {D.}~\bibnamefont {Mallick}}, \ and\ \bibinfo {author} {\bibfnamefont {B.}~\bibnamefont {Mohanty}},\ }\href {\doibase 10.1016/j.ppnp.2022.103960} {\bibfield  {journal} {\bibinfo  {journal} {Prog. Part. Nucl. Phys.}\ }\textbf {\bibinfo {volume} {125}},\ \bibinfo {pages} {103960} (\bibinfo {year} {2022})},\ \Eprint {http://arxiv.org/abs/2203.07817} {arXiv:2203.07817 [nucl-ex]} \BibitemShut {NoStop}%
\bibitem [{\citenamefont {Fukushima}\ and\ \citenamefont {Hatsuda}(2011)}]{Fukushima:2010bq}%
  \BibitemOpen
  \bibfield  {author} {\bibinfo {author} {\bibfnamefont {K.}~\bibnamefont {Fukushima}}\ and\ \bibinfo {author} {\bibfnamefont {T.}~\bibnamefont {Hatsuda}},\ }\href {\doibase 10.1088/0034-4885/74/1/014001} {\bibfield  {journal} {\bibinfo  {journal} {Rept. Prog. Phys.}\ }\textbf {\bibinfo {volume} {74}},\ \bibinfo {pages} {014001} (\bibinfo {year} {2011})},\ \Eprint {http://arxiv.org/abs/1005.4814} {arXiv:1005.4814 [hep-ph]} \BibitemShut {NoStop}%
\bibitem [{\citenamefont {Luo}\ and\ \citenamefont {Xu}(2017)}]{Luo:2017faz}%
  \BibitemOpen
  \bibfield  {author} {\bibinfo {author} {\bibfnamefont {X.}~\bibnamefont {Luo}}\ and\ \bibinfo {author} {\bibfnamefont {N.}~\bibnamefont {Xu}},\ }\href {\doibase 10.1007/s41365-017-0257-0} {\bibfield  {journal} {\bibinfo  {journal} {Nucl. Sci. Tech.}\ }\textbf {\bibinfo {volume} {28}},\ \bibinfo {pages} {112} (\bibinfo {year} {2017})},\ \Eprint {http://arxiv.org/abs/1701.02105} {arXiv:1701.02105 [nucl-ex]} \BibitemShut {NoStop}%
\bibitem [{\citenamefont {Bzdak}\ \emph {et~al.}(2020)\citenamefont {Bzdak}, \citenamefont {Esumi}, \citenamefont {Koch}, \citenamefont {Liao}, \citenamefont {Stephanov},\ and\ \citenamefont {Xu}}]{Bzdak:2019pkr}%
  \BibitemOpen
  \bibfield  {author} {\bibinfo {author} {\bibfnamefont {A.}~\bibnamefont {Bzdak}}, \bibinfo {author} {\bibfnamefont {S.}~\bibnamefont {Esumi}}, \bibinfo {author} {\bibfnamefont {V.}~\bibnamefont {Koch}}, \bibinfo {author} {\bibfnamefont {J.}~\bibnamefont {Liao}}, \bibinfo {author} {\bibfnamefont {M.}~\bibnamefont {Stephanov}}, \ and\ \bibinfo {author} {\bibfnamefont {N.}~\bibnamefont {Xu}},\ }\href {\doibase 10.1016/j.physrep.2020.01.005} {\bibfield  {journal} {\bibinfo  {journal} {Phys. Rept.}\ }\textbf {\bibinfo {volume} {853}},\ \bibinfo {pages} {1} (\bibinfo {year} {2020})},\ \Eprint {http://arxiv.org/abs/1906.00936} {arXiv:1906.00936 [nucl-th]} \BibitemShut {NoStop}%
\bibitem [{\citenamefont {Almaalol}\ \emph {et~al.}(2022)\citenamefont {Almaalol} \emph {et~al.}}]{Almaalol:2022xwv}%
  \BibitemOpen
  \bibfield  {author} {\bibinfo {author} {\bibfnamefont {D.}~\bibnamefont {Almaalol}} \emph {et~al.},\ }\href@noop {} {\  (\bibinfo {year} {2022})},\ \Eprint {http://arxiv.org/abs/2209.05009} {arXiv:2209.05009 [nucl-ex]} \BibitemShut {NoStop}%
\bibitem [{\citenamefont {Ruan}\ \emph {et~al.}(2018)\citenamefont {Ruan} \emph {et~al.}}]{Ruan:2018fpo}%
  \BibitemOpen
  \bibfield  {author} {\bibinfo {author} {\bibfnamefont {S.}~\bibnamefont {Ruan}} \emph {et~al.},\ }\href {\doibase 10.1016/j.nima.2018.02.052} {\bibfield  {journal} {\bibinfo  {journal} {Nucl. Instrum. Meth. A}\ }\textbf {\bibinfo {volume} {892}},\ \bibinfo {pages} {53} (\bibinfo {year} {2018})}\BibitemShut {NoStop}%
\bibitem [{\citenamefont {Asakawa}\ and\ \citenamefont {Kitazawa}(2016)}]{Asakawa:2015ybt}%
  \BibitemOpen
  \bibfield  {author} {\bibinfo {author} {\bibfnamefont {M.}~\bibnamefont {Asakawa}}\ and\ \bibinfo {author} {\bibfnamefont {M.}~\bibnamefont {Kitazawa}},\ }\href {\doibase 10.1016/j.ppnp.2016.04.002} {\bibfield  {journal} {\bibinfo  {journal} {Prog. Part. Nucl. Phys.}\ }\textbf {\bibinfo {volume} {90}},\ \bibinfo {pages} {299} (\bibinfo {year} {2016})},\ \Eprint {http://arxiv.org/abs/1512.05038} {arXiv:1512.05038 [nucl-th]} \BibitemShut {NoStop}%
\bibitem [{\citenamefont {Braun-Munzinger}\ \emph {et~al.}(2017)\citenamefont {Braun-Munzinger}, \citenamefont {Rustamov},\ and\ \citenamefont {Stachel}}]{Braun-Munzinger:2016yjz}%
  \BibitemOpen
  \bibfield  {author} {\bibinfo {author} {\bibfnamefont {P.}~\bibnamefont {Braun-Munzinger}}, \bibinfo {author} {\bibfnamefont {A.}~\bibnamefont {Rustamov}}, \ and\ \bibinfo {author} {\bibfnamefont {J.}~\bibnamefont {Stachel}},\ }\href {\doibase 10.1016/j.nuclphysa.2017.01.011} {\bibfield  {journal} {\bibinfo  {journal} {Nucl. Phys. A}\ }\textbf {\bibinfo {volume} {960}},\ \bibinfo {pages} {114} (\bibinfo {year} {2017})},\ \Eprint {http://arxiv.org/abs/1612.00702} {arXiv:1612.00702 [nucl-th]} \BibitemShut {NoStop}%
\bibitem [{\citenamefont {Acharya}\ \emph {et~al.}(2020)\citenamefont {Acharya} \emph {et~al.}}]{ALICE:2019nbs}%
  \BibitemOpen
  \bibfield  {author} {\bibinfo {author} {\bibfnamefont {S.}~\bibnamefont {Acharya}} \emph {et~al.} (\bibinfo {collaboration} {ALICE}),\ }\href {\doibase 10.1016/j.physletb.2020.135564} {\bibfield  {journal} {\bibinfo  {journal} {Phys. Lett. B}\ }\textbf {\bibinfo {volume} {807}},\ \bibinfo {pages} {135564} (\bibinfo {year} {2020})},\ \Eprint {http://arxiv.org/abs/1910.14396} {arXiv:1910.14396 [nucl-ex]} \BibitemShut {NoStop}%
\bibitem [{\citenamefont {Borsanyi}\ \emph {et~al.}(2012)\citenamefont {Borsanyi}, \citenamefont {Fodor}, \citenamefont {Katz}, \citenamefont {Krieg}, \citenamefont {Ratti},\ and\ \citenamefont {Szabo}}]{Borsanyi:2011sw}%
  \BibitemOpen
  \bibfield  {author} {\bibinfo {author} {\bibfnamefont {S.}~\bibnamefont {Borsanyi}}, \bibinfo {author} {\bibfnamefont {Z.}~\bibnamefont {Fodor}}, \bibinfo {author} {\bibfnamefont {S.~D.}\ \bibnamefont {Katz}}, \bibinfo {author} {\bibfnamefont {S.}~\bibnamefont {Krieg}}, \bibinfo {author} {\bibfnamefont {C.}~\bibnamefont {Ratti}}, \ and\ \bibinfo {author} {\bibfnamefont {K.}~\bibnamefont {Szabo}},\ }\href {\doibase 10.1007/JHEP01(2012)138} {\bibfield  {journal} {\bibinfo  {journal} {JHEP}\ }\textbf {\bibinfo {volume} {01}},\ \bibinfo {pages} {138} (\bibinfo {year} {2012})},\ \Eprint {http://arxiv.org/abs/1112.4416} {arXiv:1112.4416 [hep-lat]} \BibitemShut {NoStop}%
\bibitem [{\citenamefont {Bazavov}\ \emph {et~al.}(2012)\citenamefont {Bazavov} \emph {et~al.}}]{HotQCD:2012fhj}%
  \BibitemOpen
  \bibfield  {author} {\bibinfo {author} {\bibfnamefont {A.}~\bibnamefont {Bazavov}} \emph {et~al.} (\bibinfo {collaboration} {HotQCD}),\ }\href {\doibase 10.1103/PhysRevD.86.034509} {\bibfield  {journal} {\bibinfo  {journal} {Phys. Rev. D}\ }\textbf {\bibinfo {volume} {86}},\ \bibinfo {pages} {034509} (\bibinfo {year} {2012})},\ \Eprint {http://arxiv.org/abs/1203.0784} {arXiv:1203.0784 [hep-lat]} \BibitemShut {NoStop}%
\bibitem [{\citenamefont {Bazavov}\ \emph {et~al.}(2014)\citenamefont {Bazavov} \emph {et~al.}}]{Bazavov:2014xya}%
  \BibitemOpen
  \bibfield  {author} {\bibinfo {author} {\bibfnamefont {A.}~\bibnamefont {Bazavov}} \emph {et~al.},\ }\href {\doibase 10.1103/PhysRevLett.113.072001} {\bibfield  {journal} {\bibinfo  {journal} {Phys. Rev. Lett.}\ }\textbf {\bibinfo {volume} {113}},\ \bibinfo {pages} {072001} (\bibinfo {year} {2014})},\ \Eprint {http://arxiv.org/abs/1404.6511} {arXiv:1404.6511 [hep-lat]} \BibitemShut {NoStop}%
\bibitem [{\citenamefont {Stephanov}(2011)}]{Stephanov:2011pb}%
  \BibitemOpen
  \bibfield  {author} {\bibinfo {author} {\bibfnamefont {M.~A.}\ \bibnamefont {Stephanov}},\ }\href {\doibase 10.1103/PhysRevLett.107.052301} {\bibfield  {journal} {\bibinfo  {journal} {Phys. Rev. Lett.}\ }\textbf {\bibinfo {volume} {107}},\ \bibinfo {pages} {052301} (\bibinfo {year} {2011})},\ \Eprint {http://arxiv.org/abs/1104.1627} {arXiv:1104.1627 [hep-ph]} \BibitemShut {NoStop}%
\bibitem [{\citenamefont {Karsch}\ and\ \citenamefont {Redlich}(2011)}]{Karsch:2010ck}%
  \BibitemOpen
  \bibfield  {author} {\bibinfo {author} {\bibfnamefont {F.}~\bibnamefont {Karsch}}\ and\ \bibinfo {author} {\bibfnamefont {K.}~\bibnamefont {Redlich}},\ }\href {\doibase 10.1016/j.physletb.2010.10.046} {\bibfield  {journal} {\bibinfo  {journal} {Phys. Lett. B}\ }\textbf {\bibinfo {volume} {695}},\ \bibinfo {pages} {136} (\bibinfo {year} {2011})},\ \Eprint {http://arxiv.org/abs/1007.2581} {arXiv:1007.2581 [hep-ph]} \BibitemShut {NoStop}%
\bibitem [{\citenamefont {Borsanyi}\ \emph {et~al.}(2014)\citenamefont {Borsanyi}, \citenamefont {Fodor}, \citenamefont {Katz}, \citenamefont {Krieg}, \citenamefont {Ratti},\ and\ \citenamefont {Szabo}}]{Borsanyi:2014ewa}%
  \BibitemOpen
  \bibfield  {author} {\bibinfo {author} {\bibfnamefont {S.}~\bibnamefont {Borsanyi}}, \bibinfo {author} {\bibfnamefont {Z.}~\bibnamefont {Fodor}}, \bibinfo {author} {\bibfnamefont {S.~D.}\ \bibnamefont {Katz}}, \bibinfo {author} {\bibfnamefont {S.}~\bibnamefont {Krieg}}, \bibinfo {author} {\bibfnamefont {C.}~\bibnamefont {Ratti}}, \ and\ \bibinfo {author} {\bibfnamefont {K.~K.}\ \bibnamefont {Szabo}},\ }\href {\doibase 10.1103/PhysRevLett.113.052301} {\bibfield  {journal} {\bibinfo  {journal} {Phys. Rev. Lett.}\ }\textbf {\bibinfo {volume} {113}},\ \bibinfo {pages} {052301} (\bibinfo {year} {2014})},\ \Eprint {http://arxiv.org/abs/1403.4576} {arXiv:1403.4576 [hep-lat]} \BibitemShut {NoStop}%
\bibitem [{\citenamefont {Albright}\ \emph {et~al.}(2015)\citenamefont {Albright}, \citenamefont {Kapusta},\ and\ \citenamefont {Young}}]{Albright:2015uua}%
  \BibitemOpen
  \bibfield  {author} {\bibinfo {author} {\bibfnamefont {M.}~\bibnamefont {Albright}}, \bibinfo {author} {\bibfnamefont {J.}~\bibnamefont {Kapusta}}, \ and\ \bibinfo {author} {\bibfnamefont {C.}~\bibnamefont {Young}},\ }\href {\doibase 10.1103/PhysRevC.92.044904} {\bibfield  {journal} {\bibinfo  {journal} {Phys. Rev. C}\ }\textbf {\bibinfo {volume} {92}},\ \bibinfo {pages} {044904} (\bibinfo {year} {2015})},\ \Eprint {http://arxiv.org/abs/1506.03408} {arXiv:1506.03408 [nucl-th]} \BibitemShut {NoStop}%
\bibitem [{\citenamefont {Fu}\ \emph {et~al.}(2016)\citenamefont {Fu}, \citenamefont {Pawlowski}, \citenamefont {Rennecke},\ and\ \citenamefont {Schaefer}}]{Fu:2016tey}%
  \BibitemOpen
  \bibfield  {author} {\bibinfo {author} {\bibfnamefont {W.-j.}\ \bibnamefont {Fu}}, \bibinfo {author} {\bibfnamefont {J.~M.}\ \bibnamefont {Pawlowski}}, \bibinfo {author} {\bibfnamefont {F.}~\bibnamefont {Rennecke}}, \ and\ \bibinfo {author} {\bibfnamefont {B.-J.}\ \bibnamefont {Schaefer}},\ }\href {\doibase 10.1103/PhysRevD.94.116020} {\bibfield  {journal} {\bibinfo  {journal} {Phys. Rev. D}\ }\textbf {\bibinfo {volume} {94}},\ \bibinfo {pages} {116020} (\bibinfo {year} {2016})},\ \Eprint {http://arxiv.org/abs/1608.04302} {arXiv:1608.04302 [hep-ph]} \BibitemShut {NoStop}%
\bibitem [{\citenamefont {Lu}\ \emph {et~al.}(2022)\citenamefont {Lu}, \citenamefont {Gao}, \citenamefont {Luo}, \citenamefont {Chang},\ and\ \citenamefont {Liu}}]{Lu:2022nkz}%
  \BibitemOpen
  \bibfield  {author} {\bibinfo {author} {\bibfnamefont {Y.}~\bibnamefont {Lu}}, \bibinfo {author} {\bibfnamefont {F.}~\bibnamefont {Gao}}, \bibinfo {author} {\bibfnamefont {X.}~\bibnamefont {Luo}}, \bibinfo {author} {\bibfnamefont {L.}~\bibnamefont {Chang}}, \ and\ \bibinfo {author} {\bibfnamefont {Y.-x.}\ \bibnamefont {Liu}},\ }\href@noop {} {\  (\bibinfo {year} {2022})},\ \Eprint {http://arxiv.org/abs/2211.03401} {arXiv:2211.03401 [hep-ph]} \BibitemShut {NoStop}%
\bibitem [{\citenamefont {Almasi}\ \emph {et~al.}(2017)\citenamefont {Almasi}, \citenamefont {Friman},\ and\ \citenamefont {Redlich}}]{Almasi:2017bhq}%
  \BibitemOpen
  \bibfield  {author} {\bibinfo {author} {\bibfnamefont {G.~A.}\ \bibnamefont {Almasi}}, \bibinfo {author} {\bibfnamefont {B.}~\bibnamefont {Friman}}, \ and\ \bibinfo {author} {\bibfnamefont {K.}~\bibnamefont {Redlich}},\ }\href {\doibase 10.1103/PhysRevD.96.014027} {\bibfield  {journal} {\bibinfo  {journal} {Phys. Rev. D}\ }\textbf {\bibinfo {volume} {96}},\ \bibinfo {pages} {014027} (\bibinfo {year} {2017})},\ \Eprint {http://arxiv.org/abs/1703.05947} {arXiv:1703.05947 [hep-ph]} \BibitemShut {NoStop}%
\bibitem [{\citenamefont {Vovchenko}\ \emph {et~al.}(2018)\citenamefont {Vovchenko}, \citenamefont {Jiang}, \citenamefont {Gorenstein},\ and\ \citenamefont {Stoecker}}]{Vovchenko:2017ayq}%
  \BibitemOpen
  \bibfield  {author} {\bibinfo {author} {\bibfnamefont {V.}~\bibnamefont {Vovchenko}}, \bibinfo {author} {\bibfnamefont {L.}~\bibnamefont {Jiang}}, \bibinfo {author} {\bibfnamefont {M.~I.}\ \bibnamefont {Gorenstein}}, \ and\ \bibinfo {author} {\bibfnamefont {H.}~\bibnamefont {Stoecker}},\ }\href {\doibase 10.1103/PhysRevC.98.024910} {\bibfield  {journal} {\bibinfo  {journal} {Phys. Rev. C}\ }\textbf {\bibinfo {volume} {98}},\ \bibinfo {pages} {024910} (\bibinfo {year} {2018})},\ \Eprint {http://arxiv.org/abs/1711.07260} {arXiv:1711.07260 [nucl-th]} \BibitemShut {NoStop}%
\bibitem [{\citenamefont {Bellwied}\ \emph {et~al.}(2020)\citenamefont {Bellwied}, \citenamefont {Borsanyi}, \citenamefont {Fodor}, \citenamefont {Guenther}, \citenamefont {Noronha-Hostler}, \citenamefont {Parotto}, \citenamefont {Pasztor}, \citenamefont {Ratti},\ and\ \citenamefont {Stafford}}]{Bellwied:2019pxh}%
  \BibitemOpen
  \bibfield  {author} {\bibinfo {author} {\bibfnamefont {R.}~\bibnamefont {Bellwied}}, \bibinfo {author} {\bibfnamefont {S.}~\bibnamefont {Borsanyi}}, \bibinfo {author} {\bibfnamefont {Z.}~\bibnamefont {Fodor}}, \bibinfo {author} {\bibfnamefont {J.~N.}\ \bibnamefont {Guenther}}, \bibinfo {author} {\bibfnamefont {J.}~\bibnamefont {Noronha-Hostler}}, \bibinfo {author} {\bibfnamefont {P.}~\bibnamefont {Parotto}}, \bibinfo {author} {\bibfnamefont {A.}~\bibnamefont {Pasztor}}, \bibinfo {author} {\bibfnamefont {C.}~\bibnamefont {Ratti}}, \ and\ \bibinfo {author} {\bibfnamefont {J.~M.}\ \bibnamefont {Stafford}},\ }\href {\doibase 10.1103/PhysRevD.101.034506} {\bibfield  {journal} {\bibinfo  {journal} {Phys. Rev. D}\ }\textbf {\bibinfo {volume} {101}},\ \bibinfo {pages} {034506} (\bibinfo {year} {2020})},\ \Eprint {http://arxiv.org/abs/1910.14592} {arXiv:1910.14592 [hep-lat]} \BibitemShut {NoStop}%
\bibitem [{\citenamefont {Sakaida}\ \emph {et~al.}(2017)\citenamefont {Sakaida}, \citenamefont {Asakawa}, \citenamefont {Fujii},\ and\ \citenamefont {Kitazawa}}]{Sakaida:2017rtj}%
  \BibitemOpen
  \bibfield  {author} {\bibinfo {author} {\bibfnamefont {M.}~\bibnamefont {Sakaida}}, \bibinfo {author} {\bibfnamefont {M.}~\bibnamefont {Asakawa}}, \bibinfo {author} {\bibfnamefont {H.}~\bibnamefont {Fujii}}, \ and\ \bibinfo {author} {\bibfnamefont {M.}~\bibnamefont {Kitazawa}},\ }\href {\doibase 10.1103/PhysRevC.95.064905} {\bibfield  {journal} {\bibinfo  {journal} {Phys. Rev. C}\ }\textbf {\bibinfo {volume} {95}},\ \bibinfo {pages} {064905} (\bibinfo {year} {2017})},\ \Eprint {http://arxiv.org/abs/1703.08008} {arXiv:1703.08008 [nucl-th]} \BibitemShut {NoStop}%
\bibitem [{\citenamefont {Nahrgang}\ \emph {et~al.}(2019)\citenamefont {Nahrgang}, \citenamefont {Bluhm}, \citenamefont {Schaefer},\ and\ \citenamefont {Bass}}]{Nahrgang:2018afz}%
  \BibitemOpen
  \bibfield  {author} {\bibinfo {author} {\bibfnamefont {M.}~\bibnamefont {Nahrgang}}, \bibinfo {author} {\bibfnamefont {M.}~\bibnamefont {Bluhm}}, \bibinfo {author} {\bibfnamefont {T.}~\bibnamefont {Schaefer}}, \ and\ \bibinfo {author} {\bibfnamefont {S.~A.}\ \bibnamefont {Bass}},\ }\href {\doibase 10.1103/PhysRevD.99.116015} {\bibfield  {journal} {\bibinfo  {journal} {Phys. Rev. D}\ }\textbf {\bibinfo {volume} {99}},\ \bibinfo {pages} {116015} (\bibinfo {year} {2019})},\ \Eprint {http://arxiv.org/abs/1804.05728} {arXiv:1804.05728 [nucl-th]} \BibitemShut {NoStop}%
\bibitem [{\citenamefont {Oliinychenko}\ and\ \citenamefont {Koch}(2019)}]{Oliinychenko:2019zfk}%
  \BibitemOpen
  \bibfield  {author} {\bibinfo {author} {\bibfnamefont {D.}~\bibnamefont {Oliinychenko}}\ and\ \bibinfo {author} {\bibfnamefont {V.}~\bibnamefont {Koch}},\ }\href {\doibase 10.1103/PhysRevLett.123.182302} {\bibfield  {journal} {\bibinfo  {journal} {Phys. Rev. Lett.}\ }\textbf {\bibinfo {volume} {123}},\ \bibinfo {pages} {182302} (\bibinfo {year} {2019})},\ \Eprint {http://arxiv.org/abs/1902.09775} {arXiv:1902.09775 [hep-ph]} \BibitemShut {NoStop}%
\bibitem [{\citenamefont {Oliinychenko}\ \emph {et~al.}(2020)\citenamefont {Oliinychenko}, \citenamefont {Shi},\ and\ \citenamefont {Koch}}]{Oliinychenko:2020cmr}%
  \BibitemOpen
  \bibfield  {author} {\bibinfo {author} {\bibfnamefont {D.}~\bibnamefont {Oliinychenko}}, \bibinfo {author} {\bibfnamefont {S.}~\bibnamefont {Shi}}, \ and\ \bibinfo {author} {\bibfnamefont {V.}~\bibnamefont {Koch}},\ }\href {\doibase 10.1103/PhysRevC.102.034904} {\bibfield  {journal} {\bibinfo  {journal} {Phys. Rev. C}\ }\textbf {\bibinfo {volume} {102}},\ \bibinfo {pages} {034904} (\bibinfo {year} {2020})},\ \Eprint {http://arxiv.org/abs/2001.08176} {arXiv:2001.08176 [hep-ph]} \BibitemShut {NoStop}%
\bibitem [{\citenamefont {Bluhm}\ \emph {et~al.}(2020)\citenamefont {Bluhm} \emph {et~al.}}]{Bluhm:2020mpc}%
  \BibitemOpen
  \bibfield  {author} {\bibinfo {author} {\bibfnamefont {M.}~\bibnamefont {Bluhm}} \emph {et~al.},\ }\href {\doibase 10.1016/j.nuclphysa.2020.122016} {\bibfield  {journal} {\bibinfo  {journal} {Nucl. Phys. A}\ }\textbf {\bibinfo {volume} {1003}},\ \bibinfo {pages} {122016} (\bibinfo {year} {2020})},\ \Eprint {http://arxiv.org/abs/2001.08831} {arXiv:2001.08831 [nucl-th]} \BibitemShut {NoStop}%
\bibitem [{\citenamefont {Landau}\ and\ \citenamefont {Lifshitz}(2013)}]{landau2013course}%
  \BibitemOpen
  \bibfield  {author} {\bibinfo {author} {\bibfnamefont {L.~D.}\ \bibnamefont {Landau}}\ and\ \bibinfo {author} {\bibfnamefont {E.~M.}\ \bibnamefont {Lifshitz}},\ }\href@noop {} {\emph {\bibinfo {title} {Course of theoretical physics, vol.~5. Statistical Physics, part 2}}}\ (\bibinfo  {publisher} {Elsevier},\ \bibinfo {year} {2013})\BibitemShut {NoStop}%
\bibitem [{\citenamefont {Kapusta}\ \emph {et~al.}(2012)\citenamefont {Kapusta}, \citenamefont {Muller},\ and\ \citenamefont {Stephanov}}]{Kapusta:2011gt}%
  \BibitemOpen
  \bibfield  {author} {\bibinfo {author} {\bibfnamefont {J.~I.}\ \bibnamefont {Kapusta}}, \bibinfo {author} {\bibfnamefont {B.}~\bibnamefont {Muller}}, \ and\ \bibinfo {author} {\bibfnamefont {M.}~\bibnamefont {Stephanov}},\ }\href {\doibase 10.1103/PhysRevC.85.054906} {\bibfield  {journal} {\bibinfo  {journal} {Phys. Rev. C}\ }\textbf {\bibinfo {volume} {85}},\ \bibinfo {pages} {054906} (\bibinfo {year} {2012})},\ \Eprint {http://arxiv.org/abs/1112.6405} {arXiv:1112.6405 [nucl-th]} \BibitemShut {NoStop}%
\bibitem [{\citenamefont {Kapusta}\ and\ \citenamefont {Plumberg}(2018)}]{Kapusta:2017hfi}%
  \BibitemOpen
  \bibfield  {author} {\bibinfo {author} {\bibfnamefont {J.~I.}\ \bibnamefont {Kapusta}}\ and\ \bibinfo {author} {\bibfnamefont {C.}~\bibnamefont {Plumberg}},\ }\href {\doibase 10.1103/PhysRevC.97.014906} {\bibfield  {journal} {\bibinfo  {journal} {Phys. Rev. C}\ }\textbf {\bibinfo {volume} {97}},\ \bibinfo {pages} {014906} (\bibinfo {year} {2018})},\ \bibinfo {note} {[Erratum: Phys. Rev. C 102, 019901 (2020)]},\ \Eprint {http://arxiv.org/abs/1710.03329} {arXiv:1710.03329 [nucl-th]} \BibitemShut {NoStop}%
\bibitem [{\citenamefont {Singh}\ \emph {et~al.}(2019)\citenamefont {Singh}, \citenamefont {Shen}, \citenamefont {McDonald}, \citenamefont {Jeon},\ and\ \citenamefont {Gale}}]{Singh:2018dpk}%
  \BibitemOpen
  \bibfield  {author} {\bibinfo {author} {\bibfnamefont {M.}~\bibnamefont {Singh}}, \bibinfo {author} {\bibfnamefont {C.}~\bibnamefont {Shen}}, \bibinfo {author} {\bibfnamefont {S.}~\bibnamefont {McDonald}}, \bibinfo {author} {\bibfnamefont {S.}~\bibnamefont {Jeon}}, \ and\ \bibinfo {author} {\bibfnamefont {C.}~\bibnamefont {Gale}},\ }\href {\doibase 10.1016/j.nuclphysa.2018.10.061} {\bibfield  {journal} {\bibinfo  {journal} {Nucl. Phys. A}\ }\textbf {\bibinfo {volume} {982}},\ \bibinfo {pages} {319} (\bibinfo {year} {2019})},\ \Eprint {http://arxiv.org/abs/1807.05451} {arXiv:1807.05451 [nucl-th]} \BibitemShut {NoStop}%
\bibitem [{\citenamefont {Tang}\ \emph {et~al.}(2023)\citenamefont {Tang}, \citenamefont {Wu},\ and\ \citenamefont {Song}}]{Tang:2023zvj}%
  \BibitemOpen
  \bibfield  {author} {\bibinfo {author} {\bibfnamefont {S.}~\bibnamefont {Tang}}, \bibinfo {author} {\bibfnamefont {S.}~\bibnamefont {Wu}}, \ and\ \bibinfo {author} {\bibfnamefont {H.}~\bibnamefont {Song}},\ }\href {\doibase 10.1103/PhysRevC.108.034901} {\bibfield  {journal} {\bibinfo  {journal} {Phys. Rev. C}\ }\textbf {\bibinfo {volume} {108}},\ \bibinfo {pages} {034901} (\bibinfo {year} {2023})},\ \Eprint {http://arxiv.org/abs/2303.15017} {arXiv:2303.15017 [nucl-th]} \BibitemShut {NoStop}%
\bibitem [{\citenamefont {Nahrgang}\ and\ \citenamefont {Bluhm}(2020)}]{Nahrgang:2020yxm}%
  \BibitemOpen
  \bibfield  {author} {\bibinfo {author} {\bibfnamefont {M.}~\bibnamefont {Nahrgang}}\ and\ \bibinfo {author} {\bibfnamefont {M.}~\bibnamefont {Bluhm}},\ }\href {\doibase 10.1103/PhysRevD.102.094017} {\bibfield  {journal} {\bibinfo  {journal} {Phys. Rev. D}\ }\textbf {\bibinfo {volume} {102}},\ \bibinfo {pages} {094017} (\bibinfo {year} {2020})},\ \Eprint {http://arxiv.org/abs/2007.10371} {arXiv:2007.10371 [nucl-th]} \BibitemShut {NoStop}%
\bibitem [{\citenamefont {An}\ \emph {et~al.}(2021)\citenamefont {An}, \citenamefont {Basar}, \citenamefont {Stephanov},\ and\ \citenamefont {Yee}}]{An:2020vri}%
  \BibitemOpen
  \bibfield  {author} {\bibinfo {author} {\bibfnamefont {X.}~\bibnamefont {An}}, \bibinfo {author} {\bibfnamefont {G.}~\bibnamefont {Basar}}, \bibinfo {author} {\bibfnamefont {M.}~\bibnamefont {Stephanov}}, \ and\ \bibinfo {author} {\bibfnamefont {H.-U.}\ \bibnamefont {Yee}},\ }\href {\doibase 10.1103/PhysRevLett.127.072301} {\bibfield  {journal} {\bibinfo  {journal} {Phys. Rev. Lett.}\ }\textbf {\bibinfo {volume} {127}},\ \bibinfo {pages} {072301} (\bibinfo {year} {2021})},\ \Eprint {http://arxiv.org/abs/2009.10742} {arXiv:2009.10742 [hep-th]} \BibitemShut {NoStop}%
\bibitem [{\citenamefont {Chattopadhyay}\ \emph {et~al.}(2024{\natexlab{a}})\citenamefont {Chattopadhyay}, \citenamefont {Ott}, \citenamefont {Schaefer},\ and\ \citenamefont {Skokov}}]{Chattopadhyay:2024jlh}%
  \BibitemOpen
  \bibfield  {author} {\bibinfo {author} {\bibfnamefont {C.}~\bibnamefont {Chattopadhyay}}, \bibinfo {author} {\bibfnamefont {J.}~\bibnamefont {Ott}}, \bibinfo {author} {\bibfnamefont {T.}~\bibnamefont {Schaefer}}, \ and\ \bibinfo {author} {\bibfnamefont {V.~V.}\ \bibnamefont {Skokov}},\ }\href {\doibase 10.1103/PhysRevLett.133.032301} {\bibfield  {journal} {\bibinfo  {journal} {Phys. Rev. Lett.}\ }\textbf {\bibinfo {volume} {133}},\ \bibinfo {pages} {032301} (\bibinfo {year} {2024}{\natexlab{a}})},\ \Eprint {http://arxiv.org/abs/2403.10608} {arXiv:2403.10608 [nucl-th]} \BibitemShut {NoStop}%
\bibitem [{\citenamefont {Kuroki}\ \emph {et~al.}(2023)\citenamefont {Kuroki}, \citenamefont {Sakai}, \citenamefont {Murase},\ and\ \citenamefont {Hirano}}]{Kuroki:2023ebq}%
  \BibitemOpen
  \bibfield  {author} {\bibinfo {author} {\bibfnamefont {K.}~\bibnamefont {Kuroki}}, \bibinfo {author} {\bibfnamefont {A.}~\bibnamefont {Sakai}}, \bibinfo {author} {\bibfnamefont {K.}~\bibnamefont {Murase}}, \ and\ \bibinfo {author} {\bibfnamefont {T.}~\bibnamefont {Hirano}},\ }\href {\doibase 10.1016/j.physletb.2023.137958} {\bibfield  {journal} {\bibinfo  {journal} {Phys. Lett. B}\ }\textbf {\bibinfo {volume} {842}},\ \bibinfo {pages} {137958} (\bibinfo {year} {2023})},\ \Eprint {http://arxiv.org/abs/2305.01977} {arXiv:2305.01977 [nucl-th]} \BibitemShut {NoStop}%
\bibitem [{\citenamefont {Basar}(2024)}]{Basar:2024srd}%
  \BibitemOpen
  \bibfield  {author} {\bibinfo {author} {\bibfnamefont {G.}~\bibnamefont {Basar}},\ }\href@noop {} {\  (\bibinfo {year} {2024})},\ \Eprint {http://arxiv.org/abs/2410.02866} {arXiv:2410.02866 [hep-th]} \BibitemShut {NoStop}%
\bibitem [{\citenamefont {Wu}\ \emph {et~al.}(2021)\citenamefont {Wu}, \citenamefont {Shen},\ and\ \citenamefont {Song}}]{Wu:2021xgu}%
  \BibitemOpen
  \bibfield  {author} {\bibinfo {author} {\bibfnamefont {S.}~\bibnamefont {Wu}}, \bibinfo {author} {\bibfnamefont {C.}~\bibnamefont {Shen}}, \ and\ \bibinfo {author} {\bibfnamefont {H.}~\bibnamefont {Song}},\ }\href {\doibase 10.1088/0256-307X/38/8/081201} {\bibfield  {journal} {\bibinfo  {journal} {Chin. Phys. Lett.}\ }\textbf {\bibinfo {volume} {38}},\ \bibinfo {pages} {081201} (\bibinfo {year} {2021})},\ \Eprint {http://arxiv.org/abs/2104.13250} {arXiv:2104.13250 [nucl-th]} \BibitemShut {NoStop}%
\bibitem [{\citenamefont {Denicol}\ \emph {et~al.}(2018)\citenamefont {Denicol}, \citenamefont {Gale}, \citenamefont {Jeon}, \citenamefont {Monnai}, \citenamefont {Schenke},\ and\ \citenamefont {Shen}}]{Denicol:2018wdp}%
  \BibitemOpen
  \bibfield  {author} {\bibinfo {author} {\bibfnamefont {G.~S.}\ \bibnamefont {Denicol}}, \bibinfo {author} {\bibfnamefont {C.}~\bibnamefont {Gale}}, \bibinfo {author} {\bibfnamefont {S.}~\bibnamefont {Jeon}}, \bibinfo {author} {\bibfnamefont {A.}~\bibnamefont {Monnai}}, \bibinfo {author} {\bibfnamefont {B.}~\bibnamefont {Schenke}}, \ and\ \bibinfo {author} {\bibfnamefont {C.}~\bibnamefont {Shen}},\ }\href {\doibase 10.1103/PhysRevC.98.034916} {\bibfield  {journal} {\bibinfo  {journal} {Phys. Rev. C}\ }\textbf {\bibinfo {volume} {98}},\ \bibinfo {pages} {034916} (\bibinfo {year} {2018})},\ \Eprint {http://arxiv.org/abs/1804.10557} {arXiv:1804.10557 [nucl-th]} \BibitemShut {NoStop}%
\bibitem [{\citenamefont {Murase}\ and\ \citenamefont {Hirano}(2013)}]{Murase:2013tma}%
  \BibitemOpen
  \bibfield  {author} {\bibinfo {author} {\bibfnamefont {K.}~\bibnamefont {Murase}}\ and\ \bibinfo {author} {\bibfnamefont {T.}~\bibnamefont {Hirano}},\ }\href@noop {} {\  (\bibinfo {year} {2013})},\ \Eprint {http://arxiv.org/abs/1304.3243} {arXiv:1304.3243 [nucl-th]} \BibitemShut {NoStop}%
\bibitem [{\citenamefont {Murase}(2019)}]{Murase:2019cwc}%
  \BibitemOpen
  \bibfield  {author} {\bibinfo {author} {\bibfnamefont {K.}~\bibnamefont {Murase}},\ }\href {\doibase 10.1016/j.aop.2019.167969} {\bibfield  {journal} {\bibinfo  {journal} {Annals Phys.}\ }\textbf {\bibinfo {volume} {411}},\ \bibinfo {pages} {167969} (\bibinfo {year} {2019})},\ \Eprint {http://arxiv.org/abs/1904.11217} {arXiv:1904.11217 [nucl-th]} \BibitemShut {NoStop}%
\bibitem [{\citenamefont {Hammelmann}\ \emph {et~al.}(2019)\citenamefont {Hammelmann}, \citenamefont {Torres-Rincon}, \citenamefont {Rose}, \citenamefont {Greif},\ and\ \citenamefont {Elfner}}]{Hammelmann:2018ath}%
  \BibitemOpen
  \bibfield  {author} {\bibinfo {author} {\bibfnamefont {J.}~\bibnamefont {Hammelmann}}, \bibinfo {author} {\bibfnamefont {J.~M.}\ \bibnamefont {Torres-Rincon}}, \bibinfo {author} {\bibfnamefont {J.-B.}\ \bibnamefont {Rose}}, \bibinfo {author} {\bibfnamefont {M.}~\bibnamefont {Greif}}, \ and\ \bibinfo {author} {\bibfnamefont {H.}~\bibnamefont {Elfner}},\ }\href {\doibase 10.1103/PhysRevD.99.076015} {\bibfield  {journal} {\bibinfo  {journal} {Phys. Rev. D}\ }\textbf {\bibinfo {volume} {99}},\ \bibinfo {pages} {076015} (\bibinfo {year} {2019})},\ \Eprint {http://arxiv.org/abs/1810.12527} {arXiv:1810.12527 [hep-ph]} \BibitemShut {NoStop}%
\bibitem [{\citenamefont {De}\ \emph {et~al.}(2022)\citenamefont {De}, \citenamefont {Shen},\ and\ \citenamefont {Kapusta}}]{De:2022tkb}%
  \BibitemOpen
  \bibfield  {author} {\bibinfo {author} {\bibfnamefont {A.}~\bibnamefont {De}}, \bibinfo {author} {\bibfnamefont {C.}~\bibnamefont {Shen}}, \ and\ \bibinfo {author} {\bibfnamefont {J.~I.}\ \bibnamefont {Kapusta}},\ }\href {\doibase 10.1103/PhysRevC.106.054903} {\bibfield  {journal} {\bibinfo  {journal} {Phys. Rev. C}\ }\textbf {\bibinfo {volume} {106}},\ \bibinfo {pages} {054903} (\bibinfo {year} {2022})},\ \Eprint {http://arxiv.org/abs/2203.02134} {arXiv:2203.02134 [nucl-th]} \BibitemShut {NoStop}%
\bibitem [{\citenamefont {Romatschke}\ and\ \citenamefont {Romatschke}(2019)}]{Romatschke:2017ejr}%
  \BibitemOpen
  \bibfield  {author} {\bibinfo {author} {\bibfnamefont {P.}~\bibnamefont {Romatschke}}\ and\ \bibinfo {author} {\bibfnamefont {U.}~\bibnamefont {Romatschke}},\ }\href {\doibase 10.1017/9781108651998} {\emph {\bibinfo {title} {Relativistic Fluid Dynamics In and Out of Equilibrium}}},\ Cambridge Monographs on Mathematical Physics\ (\bibinfo  {publisher} {Cambridge University Press},\ \bibinfo {year} {2019})\ \Eprint {http://arxiv.org/abs/1712.05815} {arXiv:1712.05815 [nucl-th]} \BibitemShut {NoStop}%
\bibitem [{\citenamefont {Schenke}\ \emph {et~al.}(2010)\citenamefont {Schenke}, \citenamefont {Jeon},\ and\ \citenamefont {Gale}}]{Schenke:2010nt}%
  \BibitemOpen
  \bibfield  {author} {\bibinfo {author} {\bibfnamefont {B.}~\bibnamefont {Schenke}}, \bibinfo {author} {\bibfnamefont {S.}~\bibnamefont {Jeon}}, \ and\ \bibinfo {author} {\bibfnamefont {C.}~\bibnamefont {Gale}},\ }\href {\doibase 10.1103/PhysRevC.82.014903} {\bibfield  {journal} {\bibinfo  {journal} {Phys. Rev. C}\ }\textbf {\bibinfo {volume} {82}},\ \bibinfo {pages} {014903} (\bibinfo {year} {2010})},\ \Eprint {http://arxiv.org/abs/1004.1408} {arXiv:1004.1408 [hep-ph]} \BibitemShut {NoStop}%
\bibitem [{\citenamefont {Paquet}\ \emph {et~al.}(2016)\citenamefont {Paquet}, \citenamefont {Shen}, \citenamefont {Denicol}, \citenamefont {Luzum}, \citenamefont {Schenke}, \citenamefont {Jeon},\ and\ \citenamefont {Gale}}]{Paquet:2015lta}%
  \BibitemOpen
  \bibfield  {author} {\bibinfo {author} {\bibfnamefont {J.-F.}\ \bibnamefont {Paquet}}, \bibinfo {author} {\bibfnamefont {C.}~\bibnamefont {Shen}}, \bibinfo {author} {\bibfnamefont {G.~S.}\ \bibnamefont {Denicol}}, \bibinfo {author} {\bibfnamefont {M.}~\bibnamefont {Luzum}}, \bibinfo {author} {\bibfnamefont {B.}~\bibnamefont {Schenke}}, \bibinfo {author} {\bibfnamefont {S.}~\bibnamefont {Jeon}}, \ and\ \bibinfo {author} {\bibfnamefont {C.}~\bibnamefont {Gale}},\ }\href {\doibase 10.1103/PhysRevC.93.044906} {\bibfield  {journal} {\bibinfo  {journal} {Phys. Rev. C}\ }\textbf {\bibinfo {volume} {93}},\ \bibinfo {pages} {044906} (\bibinfo {year} {2016})},\ \Eprint {http://arxiv.org/abs/1509.06738} {arXiv:1509.06738 [hep-ph]} \BibitemShut {NoStop}%
\bibitem [{\citenamefont {Murase}(2015)}]{Murase:2015oie}%
  \BibitemOpen
  \bibfield  {author} {\bibinfo {author} {\bibfnamefont {K.}~\bibnamefont {Murase}},\ }\emph {\bibinfo {title} {Causal hydrodynamic fluctuations and their effects on high-energy nuclear collisions}},\ \href {\doibase 10.15083/00072981} {Ph.D. thesis},\ \bibinfo  {school} {Tokyo U.} (\bibinfo {year} {2015})\BibitemShut {NoStop}%
\bibitem [{\citenamefont {Bhambure}\ \emph {et~al.}(2024)\citenamefont {Bhambure}, \citenamefont {Singh},\ and\ \citenamefont {Teaney}}]{Bhambure:2024gnf}%
  \BibitemOpen
  \bibfield  {author} {\bibinfo {author} {\bibfnamefont {J.}~\bibnamefont {Bhambure}}, \bibinfo {author} {\bibfnamefont {R.}~\bibnamefont {Singh}}, \ and\ \bibinfo {author} {\bibfnamefont {D.}~\bibnamefont {Teaney}},\ }\href@noop {} {\  (\bibinfo {year} {2024})},\ \Eprint {http://arxiv.org/abs/2412.10306} {arXiv:2412.10306 [nucl-th]} \BibitemShut {NoStop}%
\bibitem [{\citenamefont {Chattopadhyay}\ \emph {et~al.}(2024{\natexlab{b}})\citenamefont {Chattopadhyay}, \citenamefont {Ott}, \citenamefont {Schaefer},\ and\ \citenamefont {Skokov}}]{Chattopadhyay:2024bcv}%
  \BibitemOpen
  \bibfield  {author} {\bibinfo {author} {\bibfnamefont {C.}~\bibnamefont {Chattopadhyay}}, \bibinfo {author} {\bibfnamefont {J.}~\bibnamefont {Ott}}, \bibinfo {author} {\bibfnamefont {T.}~\bibnamefont {Schaefer}}, \ and\ \bibinfo {author} {\bibfnamefont {V.~V.}\ \bibnamefont {Skokov}},\ }\href@noop {} {\  (\bibinfo {year} {2024}{\natexlab{b}})},\ \Eprint {http://arxiv.org/abs/2411.15994} {arXiv:2411.15994 [nucl-th]} \BibitemShut {NoStop}%
\bibitem [{\citenamefont {Kurganov}\ and\ \citenamefont {Tadmor}(2000)}]{Kurganov:2000ovy}%
  \BibitemOpen
  \bibfield  {author} {\bibinfo {author} {\bibfnamefont {A.}~\bibnamefont {Kurganov}}\ and\ \bibinfo {author} {\bibfnamefont {E.}~\bibnamefont {Tadmor}},\ }\href {\doibase 10.1006/jcph.2000.6459} {\bibfield  {journal} {\bibinfo  {journal} {J. Comput. Phys.}\ }\textbf {\bibinfo {volume} {160}},\ \bibinfo {pages} {241} (\bibinfo {year} {2000})}\BibitemShut {NoStop}%
\bibitem [{\citenamefont {Monnai}\ \emph {et~al.}(2021)\citenamefont {Monnai}, \citenamefont {Schenke},\ and\ \citenamefont {Shen}}]{Monnai:2021kgu}%
  \BibitemOpen
  \bibfield  {author} {\bibinfo {author} {\bibfnamefont {A.}~\bibnamefont {Monnai}}, \bibinfo {author} {\bibfnamefont {B.}~\bibnamefont {Schenke}}, \ and\ \bibinfo {author} {\bibfnamefont {C.}~\bibnamefont {Shen}},\ }\href {\doibase 10.1142/S0217751X21300076} {\bibfield  {journal} {\bibinfo  {journal} {Int. J. Mod. Phys. A}\ }\textbf {\bibinfo {volume} {36}},\ \bibinfo {pages} {2130007} (\bibinfo {year} {2021})},\ \Eprint {http://arxiv.org/abs/2101.11591} {arXiv:2101.11591 [nucl-th]} \BibitemShut {NoStop}%
\bibitem [{\citenamefont {Kapusta}\ and\ \citenamefont {Torres-Rincon}(2012)}]{Kapusta:2012zb}%
  \BibitemOpen
  \bibfield  {author} {\bibinfo {author} {\bibfnamefont {J.~I.}\ \bibnamefont {Kapusta}}\ and\ \bibinfo {author} {\bibfnamefont {J.~M.}\ \bibnamefont {Torres-Rincon}},\ }\href {\doibase 10.1103/PhysRevC.86.054911} {\bibfield  {journal} {\bibinfo  {journal} {Phys. Rev. C}\ }\textbf {\bibinfo {volume} {86}},\ \bibinfo {pages} {054911} (\bibinfo {year} {2012})},\ \Eprint {http://arxiv.org/abs/1209.0675} {arXiv:1209.0675 [nucl-th]} \BibitemShut {NoStop}%
\bibitem [{\citenamefont {Pradeep}\ \emph {et~al.}(2022)\citenamefont {Pradeep}, \citenamefont {Rajagopal}, \citenamefont {Stephanov},\ and\ \citenamefont {Yin}}]{Pradeep:2022mkf}%
  \BibitemOpen
  \bibfield  {author} {\bibinfo {author} {\bibfnamefont {M.}~\bibnamefont {Pradeep}}, \bibinfo {author} {\bibfnamefont {K.}~\bibnamefont {Rajagopal}}, \bibinfo {author} {\bibfnamefont {M.}~\bibnamefont {Stephanov}}, \ and\ \bibinfo {author} {\bibfnamefont {Y.}~\bibnamefont {Yin}},\ }\href {\doibase 10.1103/PhysRevD.106.036017} {\bibfield  {journal} {\bibinfo  {journal} {Phys. Rev. D}\ }\textbf {\bibinfo {volume} {106}},\ \bibinfo {pages} {036017} (\bibinfo {year} {2022})},\ \Eprint {http://arxiv.org/abs/2204.00639} {arXiv:2204.00639 [hep-ph]} \BibitemShut {NoStop}%
\bibitem [{\citenamefont {Aasen}\ \emph {et~al.}(2023)\citenamefont {Aasen}, \citenamefont {Floerchinger}, \citenamefont {Giacalone},\ and\ \citenamefont {Guenduez}}]{Aasen:2022cid}%
  \BibitemOpen
  \bibfield  {author} {\bibinfo {author} {\bibfnamefont {A.~S.}\ \bibnamefont {Aasen}}, \bibinfo {author} {\bibfnamefont {S.}~\bibnamefont {Floerchinger}}, \bibinfo {author} {\bibfnamefont {G.}~\bibnamefont {Giacalone}}, \ and\ \bibinfo {author} {\bibfnamefont {D.}~\bibnamefont {Guenduez}},\ }\href {\doibase 10.1103/PhysRevC.108.014904} {\bibfield  {journal} {\bibinfo  {journal} {Phys. Rev. C}\ }\textbf {\bibinfo {volume} {108}},\ \bibinfo {pages} {014904} (\bibinfo {year} {2023})},\ \Eprint {http://arxiv.org/abs/2208.04806} {arXiv:2208.04806 [nucl-th]} \BibitemShut {NoStop}%
\bibitem [{\citenamefont {Cooper}\ and\ \citenamefont {Frye}(1974)}]{Cooper:1974mv}%
  \BibitemOpen
  \bibfield  {author} {\bibinfo {author} {\bibfnamefont {F.}~\bibnamefont {Cooper}}\ and\ \bibinfo {author} {\bibfnamefont {G.}~\bibnamefont {Frye}},\ }\href {\doibase 10.1103/PhysRevD.10.186} {\bibfield  {journal} {\bibinfo  {journal} {Phys. Rev. D}\ }\textbf {\bibinfo {volume} {10}},\ \bibinfo {pages} {186} (\bibinfo {year} {1974})}\BibitemShut {NoStop}%
\bibitem [{\citenamefont {Ling}\ \emph {et~al.}(2014)\citenamefont {Ling}, \citenamefont {Springer},\ and\ \citenamefont {Stephanov}}]{Ling:2013ksb}%
  \BibitemOpen
  \bibfield  {author} {\bibinfo {author} {\bibfnamefont {B.}~\bibnamefont {Ling}}, \bibinfo {author} {\bibfnamefont {T.}~\bibnamefont {Springer}}, \ and\ \bibinfo {author} {\bibfnamefont {M.}~\bibnamefont {Stephanov}},\ }\href {\doibase 10.1103/PhysRevC.89.064901} {\bibfield  {journal} {\bibinfo  {journal} {Phys. Rev. C}\ }\textbf {\bibinfo {volume} {89}},\ \bibinfo {pages} {064901} (\bibinfo {year} {2014})},\ \Eprint {http://arxiv.org/abs/1310.6036} {arXiv:1310.6036 [nucl-th]} \BibitemShut {NoStop}%
\bibitem [{\citenamefont {Pratt}(2012)}]{Pratt:2012dz}%
  \BibitemOpen
  \bibfield  {author} {\bibinfo {author} {\bibfnamefont {S.}~\bibnamefont {Pratt}},\ }\href {\doibase 10.1103/PhysRevLett.108.212301} {\bibfield  {journal} {\bibinfo  {journal} {Phys. Rev. Lett.}\ }\textbf {\bibinfo {volume} {108}},\ \bibinfo {pages} {212301} (\bibinfo {year} {2012})},\ \Eprint {http://arxiv.org/abs/1203.4578} {arXiv:1203.4578 [nucl-th]} \BibitemShut {NoStop}%
\bibitem [{\citenamefont {De}\ \emph {et~al.}(2020)\citenamefont {De}, \citenamefont {Plumberg},\ and\ \citenamefont {Kapusta}}]{De:2020yyx}%
  \BibitemOpen
  \bibfield  {author} {\bibinfo {author} {\bibfnamefont {A.}~\bibnamefont {De}}, \bibinfo {author} {\bibfnamefont {C.}~\bibnamefont {Plumberg}}, \ and\ \bibinfo {author} {\bibfnamefont {J.~I.}\ \bibnamefont {Kapusta}},\ }\href {\doibase 10.1103/PhysRevC.102.024905} {\bibfield  {journal} {\bibinfo  {journal} {Phys. Rev. C}\ }\textbf {\bibinfo {volume} {102}},\ \bibinfo {pages} {024905} (\bibinfo {year} {2020})},\ \Eprint {http://arxiv.org/abs/2003.04878} {arXiv:2003.04878 [nucl-th]} \BibitemShut {NoStop}%
\end{thebibliography}%


\end{document}